\documentclass[a4paper,11pt]{article}
\pdfoutput=1 

\usepackage{jcappub} 

\usepackage[T1]{fontenc}
\usepackage[utf8]{inputenc}
\usepackage{slashed}
\usepackage{mathtools}
\usepackage{bm}
\usepackage{lipsum}

\DeclarePairedDelimiter{\floor}{\lfloor}{\rfloor}

	\newcommand{\gs}{g_\sigma}
	\newcommand{\gw}{g_\omega}
	\newcommand{\gr}{g_\rho}
	\newcommand{\gwr}{\Lambda_{\omega\rho}}

	\newcommand{\Qn}{\bm{Q}_n}
	\newcommand{\Qp}{{Q}_p}

\title{\boldmath Landau parameters and entrainment matrix of cold stellar matter: effect of the symmetry energy and strong magnetic fields}

\author[1]{Helena Pais,\note{Corresponding author. Email: hpais@uc.pt}}
\author[2]{Oleksii Ivanytskyi,\note{Now at Institute of Theoretical Physics, University of Wroclaw, Max Born Pl. 9, 50-204 Wroclaw, Poland.}}
\author[3]{and Constan\c ca Provid\^encia\note{Email: cp@uc.pt}}

\affiliation[]{CFisUC, Department of Physics, University of Coimbra, Rua Larga P-3004-516, Coimbra, Portugal}

\abstract{Nuclear matter properties based on a relativistic approach suitable for the description of multi-component systems are calculated. We use a set of nuclear relativistic mean-field models that satisfy acceptable nuclear matter properties and neutron star observations.  The effects of the density dependence of the symmetry energy and of the Landau quantization due to the presence of a strong external magnetic field are discussed. Properties such as the proton fraction, the Landau mass, Landau parameters and entrainment matrix, the adiabatic index and speed of sound are calculated for cold $\beta$-equilibrium matter. A large dispersion on the calculated properties is obtained at two to three times saturation density $\rho_0 $. The proton Landau mass can be  as low as one third of the vacuum nucleon mass  at 2-3$~\rho_0 $. Similar effects are obtained for the Landau parameters, in particular, the ones involving protons, where the  relative dispersion of $F^0_{pp}$ and $F^1_{pp}$ is  as high as 30\% to 50\% at 2-3$~\rho_0 $. These parameters are particularly sensitive to the symmetry energy. The effect of the magnetic field on the nuclear properties is small for fields as high as 10$^{18}$G except for a small range of densities just above the crust-core transition. Tables with the EoS, and the parameters, are provided in the Supplementary Material section.}

\begin{document}
\maketitle
\flushbottom

\section{Introduction}
\label{sec1}

Recently, multi-messenger observations of merging neutron stars (NS) provided the community with a new powerful tool to study their properties \cite{abbott,abbott2}.
In combination with traditional electromagnetic probes from neutron star cooling \cite{Yakovlev2001,Yakovlev2002,Potekhin2015},   the detection of high frequency oscillations associated to  seismic vibrations of the neutron star crust \cite{Strohmayer2006,Watts2007,Lee2008,Timokhin2008}, the thermal X-ray detection  using, for instance, the Neutron Star Interior Composition Explorer  \cite{Miller2019}
and gravitational probes \cite{Anderson2001,Anderson2003,Watts2006} from isolated pulsars, these observations can shed a light on the NS composition and exotic states of matter existing in the stellar medium.

The knowledge of the nuclear matter symmetry energy is essential to determine several properties of neutron stars, including the crust-core transition \cite{Ducoin2011,Pais2016a,Pais2016},  the neutron star radius \cite{Lattimer2007,Providencia2013,Fortin2016,Alam2016,Malik2018}, the onset of the nuclear direct Urca \cite{Carriere2003,Cavagnoli2011,Fortin2016,Fortin2020}, or  quasi-periodic oscillations
in giant flares emitted by magnetized NS \cite{Steiner2009}.  Constraints imposed on the slope of the symmetry energy from  {\it ab-initio} calculations of
pure neutron matter \cite{Hebeler2013} and astrophysical observations  are
summarized as  $40 \lesssim L \lesssim 62$ MeV \cite{Lattimer2013} or
$30 \lesssim L \lesssim 86$ MeV \cite{Li2013,Oertel2017}. However,  larger values of the symmetry energy slope seem to be compatible with the recent measurements of 
 the Lead Radius EXperiment PREX-2 \cite{PREX2}. In \cite{Reed2021}, a value of $L=(106\pm37)$~MeV has been reported for the slope of the symmetry energy.

Nuclear matter properties predicted by different relativistic mean-field properties may be characterized and compared by determining the respective Landau parameters and properties such as the speed of sound, the adiabatic index or the Landau mass. These  quantities are important to  determine, for instance, neutrino scattering and other phenomena in the interior of neutron stars and, therefore,  define the evolution of neutron stars such as cooling and the hydrodynamic behavior of the nuclear stellar  matter.

In particular, the Landau parameter $F_1$ is directly related with the entrainment matrix essential to describe the hydrodynamic behaviour of superfluid components, as proposed by \cite{Andreev1975}. Entrainment corresponds to a  momentum transfer caused by the interaction between quasi-particles. In  general, the entrainment between  $n$ superfluid components, is quantified by a  $n\times n$ symmetric matrix, whose elements are expressed through the equilibrium characteristics of matter \cite{Gusakov2009}.
Macroscopic states of correlated Cooper pairs of neutrons and protons, which are responsible for phenomena of nuclear superfluidity and superconductivity, are among them \cite{Yakovlev2001,Yakovlev2002,Lattimer2004,Page2006,Takatsuka2006}

Another possible manifestation of neutron superfluidity and proton superconductivity is related to glitches, sudden jumps of the NS rotational frequency followed by its relatively slow decrease \cite{glitches}. At the moment, it is not clear whether the stellar crust is enough to describe the glitches or whether superfluidity and/or superconductivity in the outer core are also related to that phenomenon \cite{Andersson2012,Chamel2012,Chamel2013}. This second possibility requires an adequate hydrodynamical description of the NS interiors.

Neutron stars are known as the most powerful sources of the strongest magnetic fields in the Universe. Its strength on the surface was reported to be $10^{13}-10^{15}$ G \cite{SGR/APX}. Consistent determination of the magnetic field in the stellar environment is complicated due to the non-linear nature of General Relativity. Despite that, a simple estimate with the virial theorem suggests that the central magnetic field of magnetars can reach  $\sim 10^{18}$ G \cite{Lai1991}. This result is supported by the self-consistent solution of the Einstein-Maxwell equations \cite{Bocquet1995}. At such strengths, the magnetic field significantly impacts the stellar matter properties. One of the effects is the extension of the NS crust due to the formation of a succession of stable and unstable regions caused by the opening of new Landau levels of charged particles \cite{Fang1,Fang2,Fang3}. Besides, the presence of strong magnetic field breaks the spherical symmetry of magnetars, leading to their deformation \cite{Bocquet1995,Chatterjee2015,Gomes2019}.

Furthermore, proton pairing leads to the formation of a type-II superconductor supporting the magnetic field by forming quantized electromagnetic vortices \cite{Baym1969,Muzikar1981,Mendel1991}. Superconductivity of this type will  be destroyed when the applied magnetic field exceeds some critical value. In \cite{Sinha2015}, within a given number of assumptions, a value of  $B\sim 5\times 10^{16}$ G was obtained, a magnetic field intensity that is well inside the range typical for magnetars \cite{Broderick2002,Chatterjee2015,Gomes2019}.

In this work, properties of nuclear matter in $\beta$-equilibrium are calculated within the relativistic mean-field (RMF) framework. We will be interested, in particular, in studying the effect of the density dependence of the symmetry energy.  This will be possible by considering the NL3$\omega\rho$ family consisting of a set of models that only differ on the isovector channel, see \cite{NL3,Pais2016,NL3fam}. Other parametrizations recently presented, and sharing the same framework, FSU2R and FSU2H \cite{tolos,tolos2}, TM1e \cite{TM1e}, BigApple \cite{BigApple},  will also be discussed. FSU2H and FSU2R parametrizations have been proposed as two models that describe two solar-mass stars, FSU2H including also hyperons, and  that comply with other nuclear constraints, in particular, finite nuclei properties,  constraints from kaon production and
collective flow in heavy ion collisions and from theoretical {\it ab-initio} calculations of neutron matter \cite{tolos,tolos2}. These two models are based on the FSU2 parametrization \cite{fsu2} which, within a  statistical approach, has been calibrated to reproduce the ground-state properties of finite nuclei, their monopole response, and describes two-solar mass stars. Compared with the parametrizations FSU2R and FSU2H, FSU2 has a much stiffer symmetry energy. The BigApple parametrization is based on the same energy density functional as FSU2, and has been tuned to account for a 2.6 $M_\odot$  neutron star, the possible low-mass object of the GW190814 merger. It is, however, inconsistent with  constraints obtained from heavy-ion collisions or from the tidal deformability of medium-mass stars predicted in \cite{abbott2}. The TM1e parametrization has been proposed in \cite{bao2014,TM1e}, and it has been used in supernova simulations in \cite{Sumiyoshi2019}. It is based in the  TM1 parametrization \cite{tm1}, which has been calibrated to nuclear properties and, at high densities, to results from DBHF calculations, but it has a much softer symmetry energy. For reference, we will also show results for three models with a stiff symmetry energy, NL3, TM1 and FSU2, in order to discuss the effect of the  symmetry energy, and because the PREX-2 results do not exclude a stiff behaviour of the symmetry energy \cite{PREX2,Reed2021}. 

By considering these models, we can analyze the effect of the density-dependence of the symmetry energy, which is known to be important for NS, and we discuss how the symmetry energy slope affects several properties, in particular, the proton fraction, the adiabatic index, the speed of sound, the Landau mass and Landau parameters, including the entrainment matrix, which will be calculated within the relativistic approach shown in Ref.~\cite{Gusakov2009}. All these quantities are essential in the description of the hydrodynamical behavior of neutron star matter. These properties have been discussed in previous works, but in general only models with a very hard symmetry energy were considered \cite{Caillon2001,Gusakov2009,Gusakov2014}.

We will also discuss the effect of the magnetic field on the nuclear matter properties. The merging of two NS  gives naturally  origin to magnetars due to the instigation of different types of magnetic instabilities \cite{Kiuchi2015}. These mechanisms will amplify the initial magnetic fields, and local fields with intensities above $10^{17}$ G have been reported \cite{Miret2020,Uryu2021}. Although the effects of magnetization on the EoS do not affect much NS properties, such as the mass and radius \cite{Chatterjee2015,Franzon2016,Gomes2019}, some properties such as the central nuclear matter density or proton fraction are directly influenced when NS are subjected to ultra-strong magnetic fields \cite{Chatterjee2015,Franzon2016,Bao2021}.

The paper is organized as follows. In the next section, we present all the RMF models considered, and discuss their properties. Section \ref{sec3} is devoted to the description of the magnetized nuclear matter at zero temperature, while the calculation and analysis of the nuclear matter properties, Landau parameters and entrainment matrix coefficients are given in Section \ref{sec4}. Our conclusions are formulated in Section \ref{sec5}.

\section{Relativistic mean-field models}
\label{sec2}

We consider a conservative picture of stellar matter, which includes only nucleonic degrees of freedom neutralized by electrons. To model the stellar interior, we use models that include a non-linear $\omega$-meson term introduced in \cite{tm1} in order to soften the high-density behavior of the EoS. Isoscalar attractive and repulsive forces between neutrons and protons are mediated by scalar $\sigma$ and vector $\omega$ mesons, respectively. Isovector repulsion is generated by the vector $\rho$ meson. The coupling of the isoscalar-vector and isovector-vector mesons is also considered in the present models \cite{NL3fam,Pais2016}. This term allows to model the density-dependence of the symmetry energy.
The Lagrangian density of the system under consideration, where strong magnetic fields are present in a frozen-field configuration, is given by
\begin{eqnarray}
\label{I}
\mathcal{L}=\sum_{i=n,p}\mathcal{L}_i+\mathcal{L}_e+\mathcal{L}_A+\mathcal{L}_\sigma+\mathcal{L}_\omega+\mathcal{L}_\rho+\mathcal{L}_{\omega\rho} \, .
\end{eqnarray}
The first term in this expression corresponds to the nucleons:
\begin{eqnarray}
\label{II}
\mathcal{L}_i=\overline{\psi}_i
\left(i\slashed D-m^*_N-
\frac{1}{2}\mu_N\varkappa_i\sigma_{\mu\nu}F^{\mu\nu}\right)\psi_i \, .
\end{eqnarray}
For simplicity, we take the bare nucleon mass as $m_p=m_n=m_N=939$ MeV. Hereafter we work in the natural system of units, with $c=\hbar=1$. In this system of units, the electromagnetic coupling constant is $e=\sqrt{\frac{4\pi}{137}}$. The Dirac effective mass of the nucleons, and the covariant derivative,  entering in Eq.~(\ref{II}), are given, respectively, by
\begin{eqnarray}
\label{III}
m_N^*&=&m_N-g_\sigma\sigma \, ,\\
\label{IV}
iD^\mu&=&i\partial^\mu-g_\omega\omega^\mu-
\frac{g_\rho}{2}\boldsymbol\tau\cdot\boldsymbol b^\mu-
eA^\mu\frac{1+\tau_3}{2} \, .
\end{eqnarray}
$g_\sigma$, $g_\omega$ and $g_\rho$ are the nucleon-meson couplings, while the vector $\boldsymbol\tau$ stands for the Pauli matrices. The four-potential of the electromagnetic field is represented by $A^\mu$. The last term in Eq. (\ref{II}) accounts for the nucleon anomalous magnetic moment (AMM) coupled to the electromagnetic field tensor $F^{\mu\nu}=\partial^\mu A^\nu-\partial^\nu A^\mu$ through $\sigma_{\mu\nu}=\frac{i}{2}[\gamma_\mu,\gamma_\nu]$, with $\varkappa_n=-1.91315$ for the neutron and $\varkappa_p= 1.792855$ for the proton. The nuclear magneton is given by $\mu_N={e}/{2m_N}$.

The electrons contribute to the Lagrangian density as
\begin{eqnarray}
\label{V}
\mathcal{L}_e=\overline{\psi}_e(i\slashed\partial+e
\slashed{A}-m_e)\psi_e,
\end{eqnarray} 
with $m_e=0.511$ MeV the electron mass. 
Contrary to the nucleons, the effect of the leptons AMM is ignored since it is expected to be very small, even under strong magnetic fields \cite{Duncan2000}. The electromagnetic Lagrangian reads
$\mathcal{L}_A=-\frac{1}{4}F_{\mu\nu}F^{\mu\nu} \, ,$
and the mesonic kinetic, mass and interaction terms are presented in the following Lagrangian densities:
\begin{eqnarray}
\label{VII}
\mathcal{L}_\sigma&=&\frac{1}{2}\left(\partial_\mu\sigma\partial^\mu\sigma-m_\sigma^2\sigma^2\right)-\frac{1}{3!}\kappa\sigma^3-
\frac{1}{4!}\lambda\sigma^4,\\
\label{VIII}
\mathcal{L}_\omega&=&-\frac{1}{4}\Omega_{\mu\nu}\Omega^{\mu\nu}+\frac{1}{2}m_\omega^2\omega_\mu\omega^\mu+ \frac{\zeta}{4!} g_\omega^4 (\omega_{\mu}\omega^{\mu})^2,\\
\label{IX}
\mathcal{L}_\rho&=&-\frac{1}{4}{\bf R_{\mu\nu}R^{\mu\nu}}+\frac{1}{2}m_\rho^2\boldsymbol b_\mu\cdot\boldsymbol b^\mu,\\
\label{X}
\mathcal{L}_{\omega\rho}&=&\Lambda_{\omega\rho}g_\omega^2g_\rho^2\omega_\mu\omega^\mu\boldsymbol b_\mu\cdot\boldsymbol b^\mu,
\end{eqnarray}
with $m_\sigma$, $m_\omega$, and $m_\rho$ the masses of the mesons, and $\Omega^{\mu\nu}=\partial^\mu\omega^\nu-\partial^\nu\omega^\mu$ and ${\bf R}^{\mu\nu}=\partial^\mu\boldsymbol b^\nu-\partial^\nu\boldsymbol b^\mu-g_\rho\left(\boldsymbol b^\mu\times\boldsymbol b^\nu\right)$. The term $\mathcal{L}_\sigma$, that corresponds to the scalar meson $\sigma$, includes the self-interacting coupling constants $\kappa$ and $\lambda$. This isoscalar-scalar term tunes the incompressibility and effective mass of the nucleons at the nuclear saturation density $\rho_0$. The non-linear self-interacting  $\omega$-meson term, controlled by the $\zeta$ parameter, allows to soften the high-density behavior of the EoS. This term is not present in the NL3$\omega\rho$ family, including the NL3 model.
The non-linear term $\mathcal{L}_{\omega\rho}$, previously introduced in Refs.~\cite{NL3fam,Pais2016},  mixes the $\omega$ and $\rho$ mesons via the self-interacting coupling constant $\Lambda_{\omega\rho}$ and is crucial to model the density dependence of the symmetry energy. This term is absent in the NL3 and TM1 models.

Under the mean-field approximation, only temporal components of the vector meson fields attain constant and finite mean values. From this point on, we refer to these mean values as $\omega_0$ and $b_0$, while $\sigma$ stands for the scalar meson mean value. The effective chemical potentials, i.e. the Landau effective masses, in the following just referred as effective masses, of the nucleons and electrons are written respectively as
\begin{eqnarray}
\label{XIV}
\mu_i^*&=&\mu_i-g_\omega\omega_0-g_\rho t_{i3} b_0 \, , \quad i=n,p \, ,\\
\mu_e^*&=&\mu_e \, .
\end{eqnarray}
The physical chemical potentials of the nucleons, $\mu_i$, are given in terms of the baryonic $\mu_B$ and electric $\mu_Q$ ones, $\mu_i=\mu_B+Q_i\mu_Q$, with $Q_i$  being the electric charge. $t_{i3}$ is the third isospin projection of nucleon $i$ ($+\frac{1}{2}$ for protons and $-\frac{1}{2}$ for neutrons).  
The condition of $\beta$-equilibrium for neutrino-free matter is given by $\mu_n=\mu_p+\mu_{e}$ with the electronic chemical potential  $\mu_e=-\mu_Q$.
Note that $\mu_B$ is unambiguously defined by the value of the baryonic density $\rho$, while the proper value of $\mu_Q$ can be found by imposing the condition of electric neutrality, $\rho_e=\rho_p$.

As previously mentioned, in this work, we use non-linear RMF models. Besides the NL3$\omega\rho$ family, we also consider the models FSU2, FSU2R, FSU2H, BigApple, TM1e and TM1 as indicated above. The properties of these models are given in Table~\ref{table1a}.
The NL3 model \cite{NL3}, which is the head of the NL3$\omega\rho$ family, was fitted to the ground-state properties of both stable and unstable nuclei. It gives maximum masses above 2 $M_\odot$, but it has a very high slope of the symmetry energy, making this EoS very stiff above saturation density. In order to make this EoS softer, a family was built, by varying its isovector terms, but keeping fixed the isoscalar ones. In Ref.~\cite{Pais2016}, some members of this family  proved to be consistent with microscopic state-of-the-art neutron matter calculations \cite{Gandolfi2012,Hebeler2013}, in particular, the model with the slope of the symmetry energy equal to 55 MeV, NL3$\omega\rho55$.

\begin{table}[htb]
\centering
\vspace{0.5cm}
\begin{tabular}{|c|ccccc|}
\hline
 Model    &  $\rho_0$ & $B/A$ &  $K$ & $J$ & $L$ \\
\hline
NL3        &  0.148   & -16.2  & 272  & 37.3  &  118 \\
NL3$\omega\rho$88  &  0.148  & -16.2 &  272  & 34.9  &  88 \\
NL3$\omega\rho$55  &  0.148   & -16.2 &  272  & 31.7 & 55 \\
FSU2 &  0.15  & -16.3 & 238  &  37.6  & 113 \\
FSU2R & 0.15  & -16.3  & 238  &  30.7  & 47 \\
FSU2H & 0.15  & -16.3  & 238  &  30.5  & 45  \\
BigApple  & 0.155  & -16.3  & 227 &  31.3  & 40  \\ 
TM1    & 0.145 & -16.3   & 281    &  36.9  & 111  \\
TM1e    & 0.145 & -16.3   & 281    &  31.4  & 40 \\
\hline
\end{tabular}
\caption{\label{table1a}
The symmetric nuclear matter properties at saturation density for all the models considered in this work: the nuclear saturation density $\rho_0$, the binding energy per particle $B/A$, the incompressibility $K$, the symmetry energy $J$, and the slope of the symmetry energy $L$. All quantities are in MeV, except for $\rho_0$, given in fm$^{-3}$.}
\end{table}

\begin{figure*}
\centering
\includegraphics[width=1.\linewidth]{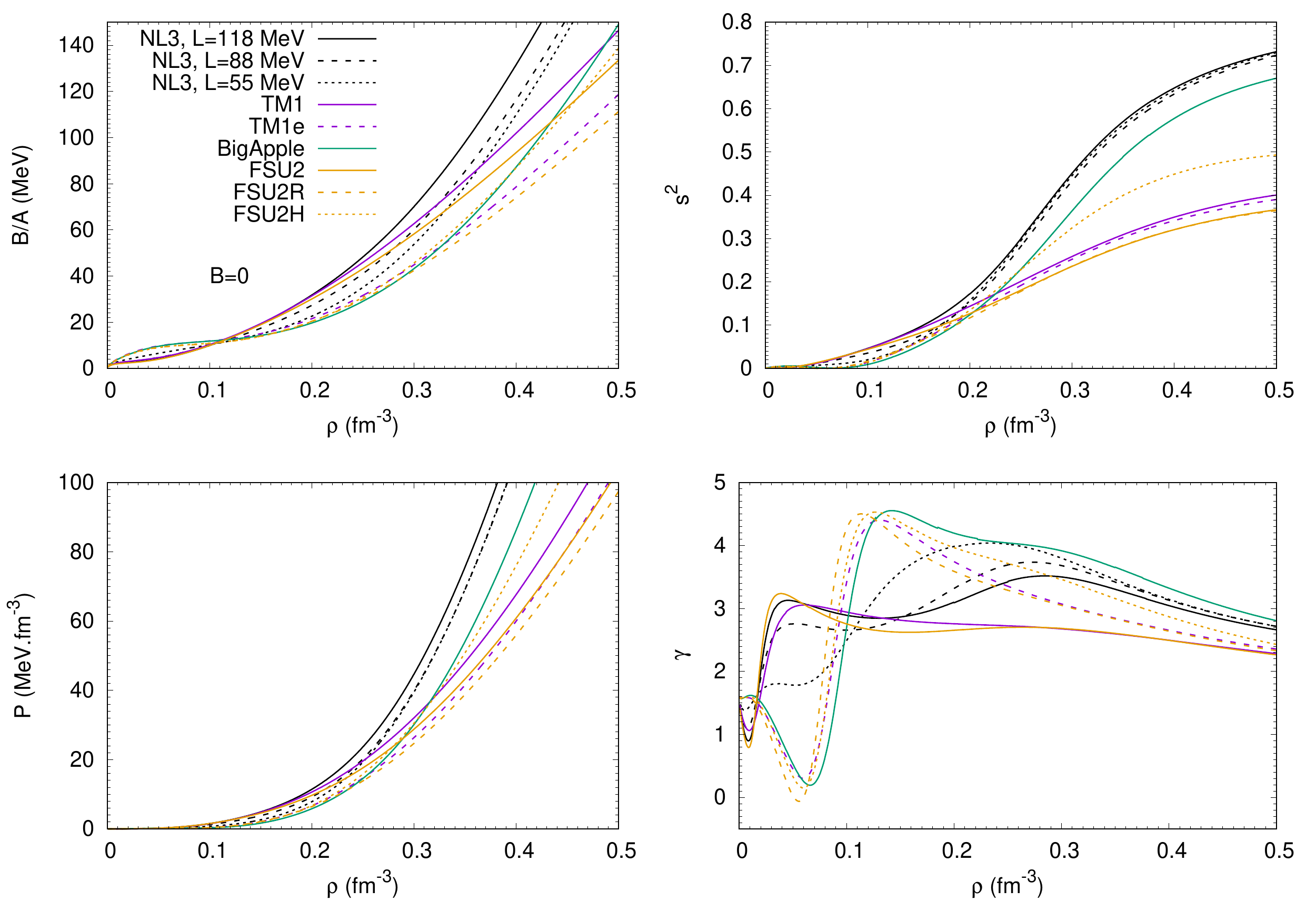} 
\caption{Binding energy per particle (top, left), pressure (bottom, left), the square of the speed of sound (top, right), and the adiabatic index (bottom, right) as a function of the baryonic density $\rho$ for cold $\beta$-equilibrium matter without magnetic field taking the models considered in this work.}
\label{fig1}
\end{figure*}

The construction of the above models requires a special comment. As said before, NL3 has a very high slope of the symmetry energy, $L=118$ MeV, not in agreement with present theoretical {\it ab-initio} calculations \cite{Gandolfi2012,Hebeler2013}. The fits of nuclear masses to experimental ones, combined with other experimental information from neutron skins, heavy ion collisions, giant dipole resonances and dipole polarizabilities, yield the symmetry energy in the range of $J=30.9\pm1.9$ MeV, and the slope of the symmetry energy in $L=51.2\pm10.7$ MeV \cite{Lattimer2013}. Another survey of a rich collection of data from terrestrial experiments and astrophysical observations yields $J=31.7\pm3.2$ MeV and $L=58.7\pm28.1$ MeV \cite{Li2013,Oertel2017}. This value of $J$ is supported by very recent Bayesian analyses performed within chiral effective field theory \cite{Drischler2020}, observational data from GW170817 and NICER \cite{Xie2019}, and experimental data on the nuclear thickness \cite{Xu2020}. For the symmetry energy slope, these Bayesian analyses suggest $L=59.8\pm4.1$ MeV, $L=66^{+12}_{-20}$ MeV and $L=41.9^{+24.6}_{-17.5}$ MeV, respectively. Note that the errors are given at $1\sigma$ confidence interval.
 
In the present analysis, we take two parametrizations of the NL3$\omega\rho$ family from Ref.~\cite{Pais2016}, one with $L=55$ MeV, that falls well into the above ranges, and the other with $L=88$ MeV, which is on the margin of the allowed interval. We label them  NL3$\omega\rho$55 and NL3$\omega\rho$88, respectively. Since the NL3$\omega\rho$55 model is consistent with a credible value of $J$ and yields the most central value of $L$, we consider it as the reference one. 

\begin{figure}[!]
\centering
\includegraphics[width=0.5\columnwidth]{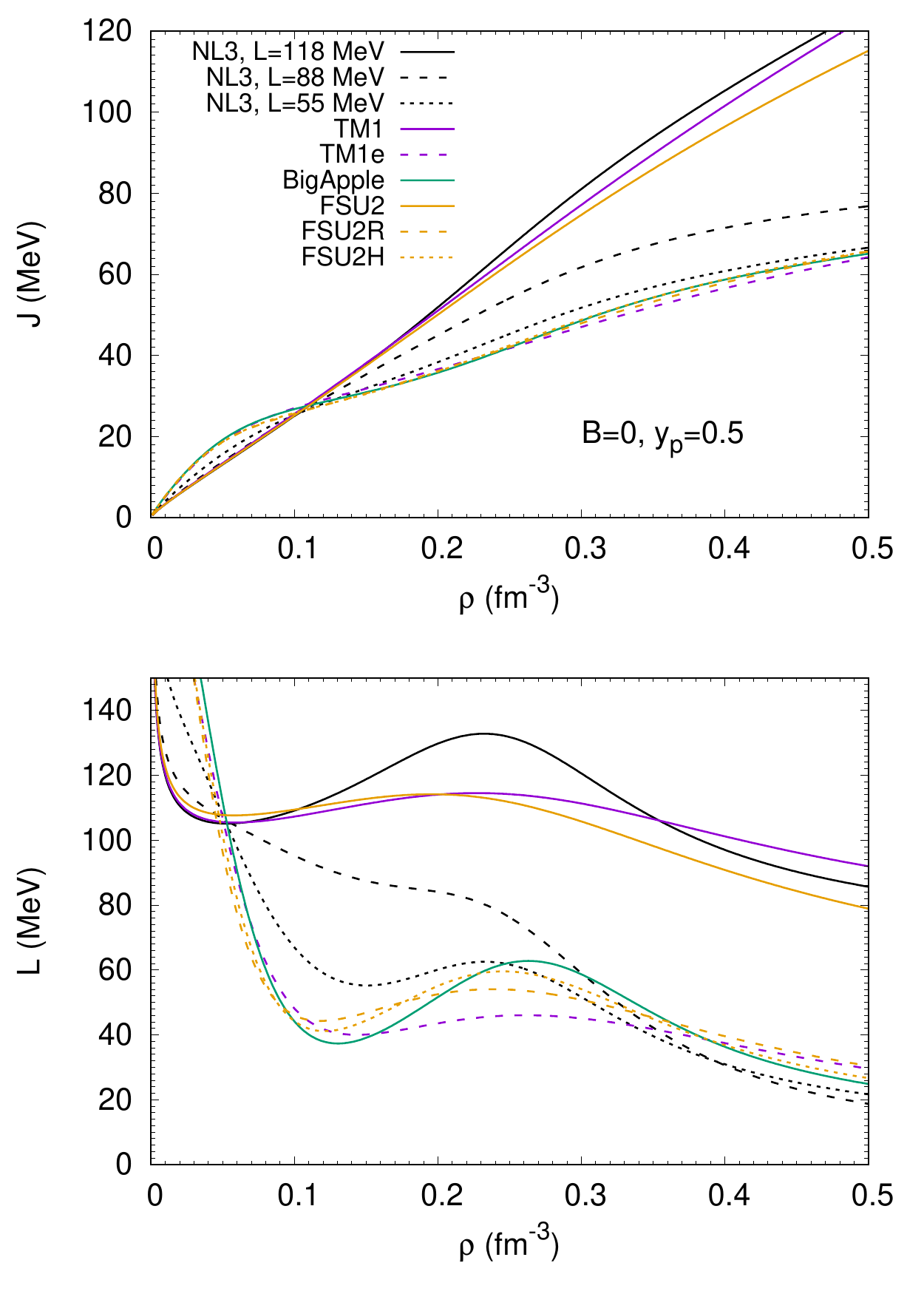}
 \caption{Symmetry energy $J$ (top) and its slope $L$ as function of the  baryonic density $\rho$ for cold symmetric nuclear matter without magnetic field and all the models considered in this work. }
\label{fig2}
\end{figure}

The EoSs of the models considered are calculated within the standard mean-field approximation. In order to analyze the effects of the symmetry energy, we plot in Fig.~\ref{fig1} the pressure $P$, the binding energy per nucleon $B/A$, the squared speed of sound
\begin{equation}
s^2=\delta P/\delta \epsilon \, , 
\end{equation}
 and the adiabatic index 
\begin{equation}
 \gamma=s^2(P+\epsilon)/P   
\end{equation}
of cold $\beta-$equilibrium nuclear matter in the absence of magnetic field as function of baryonic density, $\rho$. In these expressions, $\epsilon$ is the energy density and the derivatives are taken under the condition of $\beta$-equilibrium. It is seen that the coupling of the $\omega$ and $\rho$ mesons, controlled by the parameter $\Lambda_{\omega\rho}$, indeed, softens the EoS. Looking at Fig.~\ref{fig2}, where the symmetry energy and the correspondent slope are plotted as a function of the baryonic density for each model, we note that the three NL3$\omega\rho$ models have the same symmetry energy at 0.1 fm$^{-3}$ and, therefore, the curves of the energy per particle cross at that density. Below this density, the models with the smallest value of $L$ at saturation have the largest symmetry energy, and this is directly reflected in the energy per particle, shown in Fig.~\ref{fig1}. This same effect is present in all the other models we have considered,  see Fig.~\ref{fig2}. The effect of the non-linear $\omega$-meson term, controlled by the $\zeta$ coupling, is clearly seen in the squared speed of sound behaviour, shown in the top right panel of Fig. \ref{fig1}: models without this term, such as NL3 and NL3$\omega\rho55$, or with a very small $\zeta$ coupling constant, such as BigApple and FSU2H, become very stiff at large densities. The behaviour of this quantity, together with the one of the adiabatic index, shown in the bottom right panel of Fig.~\ref{fig1}, also reflects the density dependence of the  symmetry energy, as it easily seen from the analysis of Fig.~\ref{fig2}. Considering the NL3$\omega\rho$ family, it is seen that above  (below) saturation density, $J$ decreases (increases), the smaller the symmetry energy slope. This same behavior is reflected in the pair TM1, TM1e and FSU2 compared with FSU2R, FSU2H and BigApple. The adiabatic index is particularly sensitive to the density dependence of the symmetry energy: the softening of the symmetry energy below 0.1 fm$^{-3}$ in the models with a small $L$ has a strong effect in the adiabatic index $\gamma$. It is only above $0.15-0.25$ fm$^{-3}$, when $3\lesssim \gamma\lesssim 4$, that all models show a monotonic decreasing behavior.

The  symmetry energy influences   the fractions of neutrons and protons, which directly impact  several nuclear matter properties including the effective masses, Landau parameters and the entrainment matrix coefficients. As it will be shown in Section \ref{sec4}, some of these quantities  are indeed very sensitive to the value of $J$. 

\begin{figure}[!]
\centering
 \includegraphics[width=0.5\columnwidth]{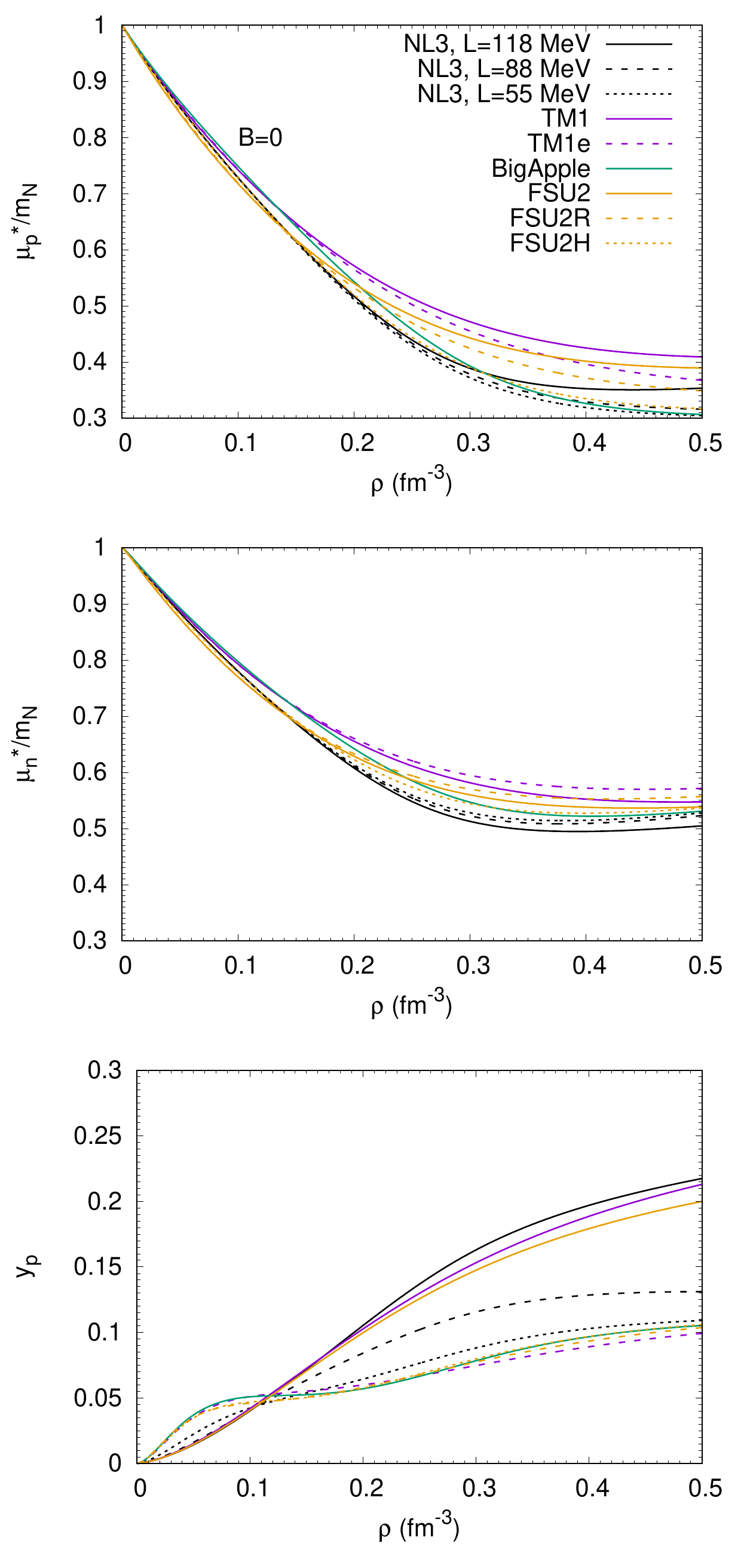}
 \caption{Normalized Landau effective masses for protons $\mu_p^*/m_N$ (top), and for neutrons $\mu_n^*/m_N$ (middle), and the proton fraction $y_p$ (bottom) as function of the  baryonic density $\rho$ for cold $\beta$-equilibrium matter without magnetic field for all the models considered in this work.}
\label{fig2a}
\end{figure}

In Fig. \ref{fig2a}, the proton (top panel) and  neutron (middle panel) effective masses are plotted together with the proton fraction  (bottom panel) of $\beta$-equilibrium matter.  Analysing the proton fraction as a function of the density, three sets of models are clearly identified: (i) TM1, NL3 and FSU2 with a proton fraction of the order of 0.2  at three times saturation density, (ii) NL3$\omega\rho$55, FSU2R, FSU2H and BigApple with a proton fraction slightly above 0.1, and (iii) NL3$\omega\rho$88 with a proton fraction close to 0.13.  The effective masses are affected by both the Dirac masses, that depend on the $\sigma$ field and coupling constant, and on the particle Fermi momenta, defined by the $\beta$-equilibrium condition.
In general, the effective masses for neutrons (protons) is smaller (larger) for models of  set (i) when compared with the models of set (ii)  or (iii) of the same family.
While at saturation density, all models have a similar effective mass, of the order of 0.7$m_N$ for neutrons and 0.65$m_N$ for protons,  at 2$\rho_0 $ (3$\rho_0 $) 
the dispersion is much larger with protons having $0.37\lesssim \mu^*_p/m_N\lesssim 0.47$  ($0.3\lesssim \mu^*_p/m_N\lesssim 0.4$), and neutrons $0.53\lesssim \mu^*_n/m_N\lesssim 0.6$ ($0.5\lesssim \mu^*_n/m_N\lesssim 0.57$). Comparing two models of the same family, that only differ in the symmetry energy, the effect on the  proton effective mass
at 3$\rho_0 $ corresponds to a decrease of the order of 11\% to 14\%  in models with a smaller slope $L$, while for the neutrons the effect is on the opposite direction and half the magnitude.

%

\section{Magnetized nuclear matter}
\label{sec3}

In the this section, we present explicit expressions for various thermodynamic quantities in the presence of a magnetic field, which, however, recover the corresponding results at $B=0$.
Since we consider an external magnetic field, only its frozen configurations are of interest for our study. We choose the reference frame in which the magnetic field of strength $B$ is aligned along the $z$-axis, i.e. ${\bf B}=(0,0,B)$. Up to an insignificant gauge transformation, this corresponds to $A^\mu=(0,0,Bx,0)$. In what follows, $B$ is given in units of the electron critical magnetic field, $B_c={m_e^2}/{e}\simeq4.4\times 10^{13}$ G, and it is quantified by $B^*=\frac{B}{B_c}$.

A constant magnetic background leads to the Landau quantization of the transverse momenta of charged particles. The transversal motion of these particles corresponds to orbiting in the plane perpendicular to ${\bf B}$. The corresponding energy levels are twice degenerated, except the lowest level, whose degeneracy is one. Another effect of $B$ is related to shifting the energy levels due to the presence of AMM. Thus, the single-particle energies of nucleons and electrons are given by
\begin{eqnarray}
\label{XI}
\varepsilon_n&=&\sqrt{k_{||}^2+
\left(\sqrt{k_\perp^2+{m_N^*}^2}-s\mu_N\kappa_nB\right)^2} \, ,\\
\label{XII}
\varepsilon_p&=&\sqrt{k_{||}^2+
\left(\sqrt{2\nu eB+{m_N^*}^2}-s\mu_N\kappa_pB\right)^2} \, ,\\
\label{XIII}
\varepsilon_e &=&\sqrt{k_{||}^2+2\nu eB+m_e^2} \, ,
\end{eqnarray}
where $k_{||}$ and $k_\perp$ are the moment components parallel and perpendicular to the magnetic field, respectively. The index $s=\pm1$ labels spin states and $\nu=l+\frac{1}{2}(1-s)$ with $l=0,1,2,...$ is the quantum number corresponding to the Landau levels. At zero temperature, nuclear matter exists in the lowest energy state. Therefore, only the single particle states with $\varepsilon_i\le\mu_i^*$ contribute to the system pressure, given by
\begin{eqnarray}
\label{XV}
P=\sum_{i=n,p,e}P_i+P_m \, ,
\end{eqnarray}
where the contributions of nucleons, electrons and mesons are given respectively by
\begin{eqnarray}
\label{XVI}
P_n&=&\sum_s\int\frac{d\bf k}{(2\pi)^3}~
(\mu_n^*-\varepsilon_n)\theta(\mu_n^*-\varepsilon_n) \, , \\
\label{XVII}
P_i&=&\sum_\nu g_\nu eB\int\frac{dk_{||}}{(2\pi)^2}
(\mu_i^*-\varepsilon_i)\theta(\mu_i^*-\varepsilon_i) \, ,
 \\
\label{XVIII}
P_m&=&-\frac{m_\sigma^2}{2}\sigma^2-\frac{\kappa}{3!}\sigma^3-\frac{\lambda}{4!}\sigma^4 +\frac{m_\omega^2}{2}\omega_0^2 + \frac{\zeta g_\omega^4}{4!}\omega_0^4 \\ \nonumber
&+&\frac{m_\rho}{2}b_0^2+
\Lambda_{\omega\rho}g_\omega^2g_\rho^2\omega_0^2 b_0^2 \, ,
\end{eqnarray}
with $i=p,e$ in Eq.~(\ref{XVII}).
The $\theta$-function, the unit step function, insures that the summation and integration are performed only over the states whose momenta and/or Landau quantum number do not exceed the Fermi ones, and/or the maximal value $\nu^{\max}$, respectively. The degeneracy factor is given by $g_\nu=1$ for $\nu=1$, and $g_\nu=2$ for $\nu\ge2$. Since spherical symmetry is broken due to the presence of the magnetic field, the Fermi surface of neutrons is not a sphere, but an ellipsoid of revolution given by
\begin{eqnarray}
\label{XIX}
k^F_{||n}(k_\perp)=\sqrt{\mu_n^{*2}-
\left(\sqrt{k_\perp^2+{m_N^*}^2}-s\mu_N\kappa_nB\right)^2}, \, 
\end{eqnarray}
where the transverse momentum of neutrons is limited by the maximal value
\begin{eqnarray}
\label{XX}
k^F_{\perp n}=\sqrt{(\mu_n^*+s\mu_N\kappa_nB)^2-{m_N^*}^2} \, .
\end{eqnarray}
For charged protons and electrons, the Fermi surface becomes even more complex due to the quantization of their transverse momenta. At a given value of the Landau quantum number, their longitudinal momenta do not exceed
\begin{eqnarray}
\label{XXI}
k^F_{||p}(\nu)&=&\sqrt{\mu_p^{*2}-
\left(\sqrt{2\nu eB+{m_N^*}^2}-s\mu_N\kappa_pB\right)^2} \, ,
\quad\\
\label{XXII}
k^F_{|| e}(\nu)&=&\sqrt{\mu_{e}^2-2\nu eB+m_e^2} \, ,
\end{eqnarray}
with $\nu$ being limited by
\begin{eqnarray}
\label{XXIII}
\nu_p^{\max}&=&\floor*{\frac{\sqrt{(\mu_p^*+s\mu_N\kappa_pB)^2-{m_N^*}^2}}{2eB}} \, ,\\
\label{XXIV}
\nu_e^{\max}&=&\floor*{\frac{\sqrt{\mu_{e}^{*2}-m_{e}^2}}{2eB}} \, .
\end{eqnarray}
Here $\floor{~}$ represents an integer part from below. 

Physical values of the mesonic field are found as solutions of the corresponding Euler-Lagrange equations. These solutions maximize the pressure, Eq.~(\ref{XV}). Number densities of nucleons and electrons are found as partial derivatives of $P$ with respect to corresponding physical chemical potentials. The baryonic density is then defined as $\rho=\rho_n+\rho_p$, and the  condition of electric neutrality must also be imposed, $\rho_p=\rho_e$. 
The energy density is given by the standard thermodynamic identity $\epsilon={\mu_B \rho}-P$. The squared speed of sound  $s^2$ and the adiabatic index  $\gamma$ were defined in the previous section. In the Appendix, for the readers' convenience,  we present the Euler-Lagrange equations for the mesonic fields along with the explicit expressions for the discussed thermodynamic quantities.

Comparing Eqs. (\ref{XI}) and (\ref{XII}), we conclude that $2\nu eB$ corresponds to the value of the squared transverse momentum of charged particles. Therefore, discrete variation of this squared momentum, i.e. $2eB$, should be associated to the continuous one, $dk_\perp^2$. In the cylindrical coordinates, this discrete variation corresponds to the phase space element $2\pi eBdk_{||}$. The latter is equivalent to the integration element $d\bf k$. This equivalence implies that, at the vanishing magnetic field, summation over $\nu$ and integration over $k_{||}$ as $\sum\limits_\nu eB\int {dk_{||}}/{(2\pi)^2}$ converges to the integration over the three-momentum as $\int{d\bf k}/{(2\pi)^3}$. Consequently, partial pressures and particle densities of charged particles given by Eqs.~(\ref{XVI}), (\ref{XVII}) and Eqs.~(\ref{A12}), (\ref{A13}) recover the standard form at $B=0$.

\begin{figure*}[!]
\centering
\begin{tabular}{cc}
\includegraphics[width=0.5\textwidth]{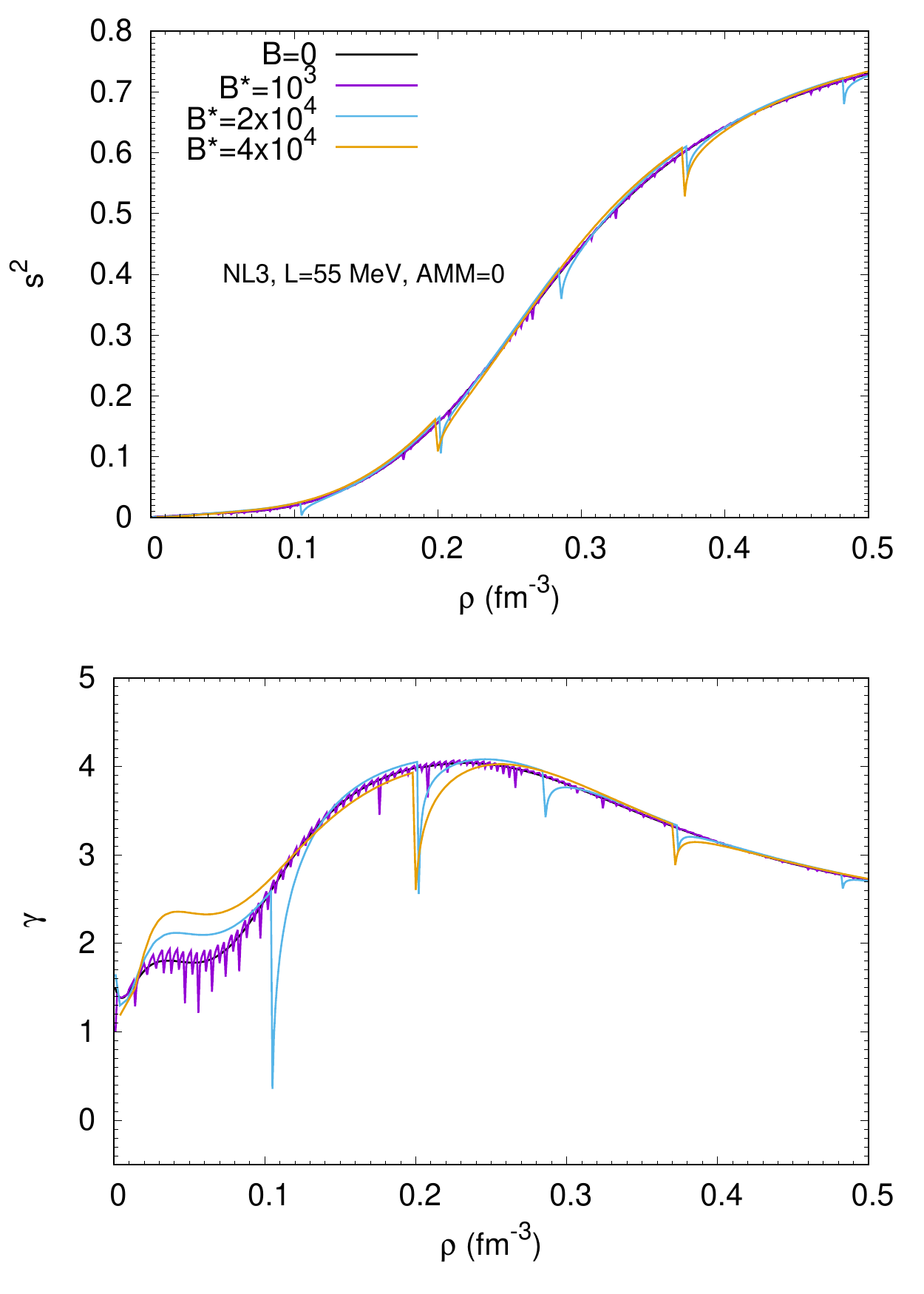} & \includegraphics[width=0.5\textwidth]{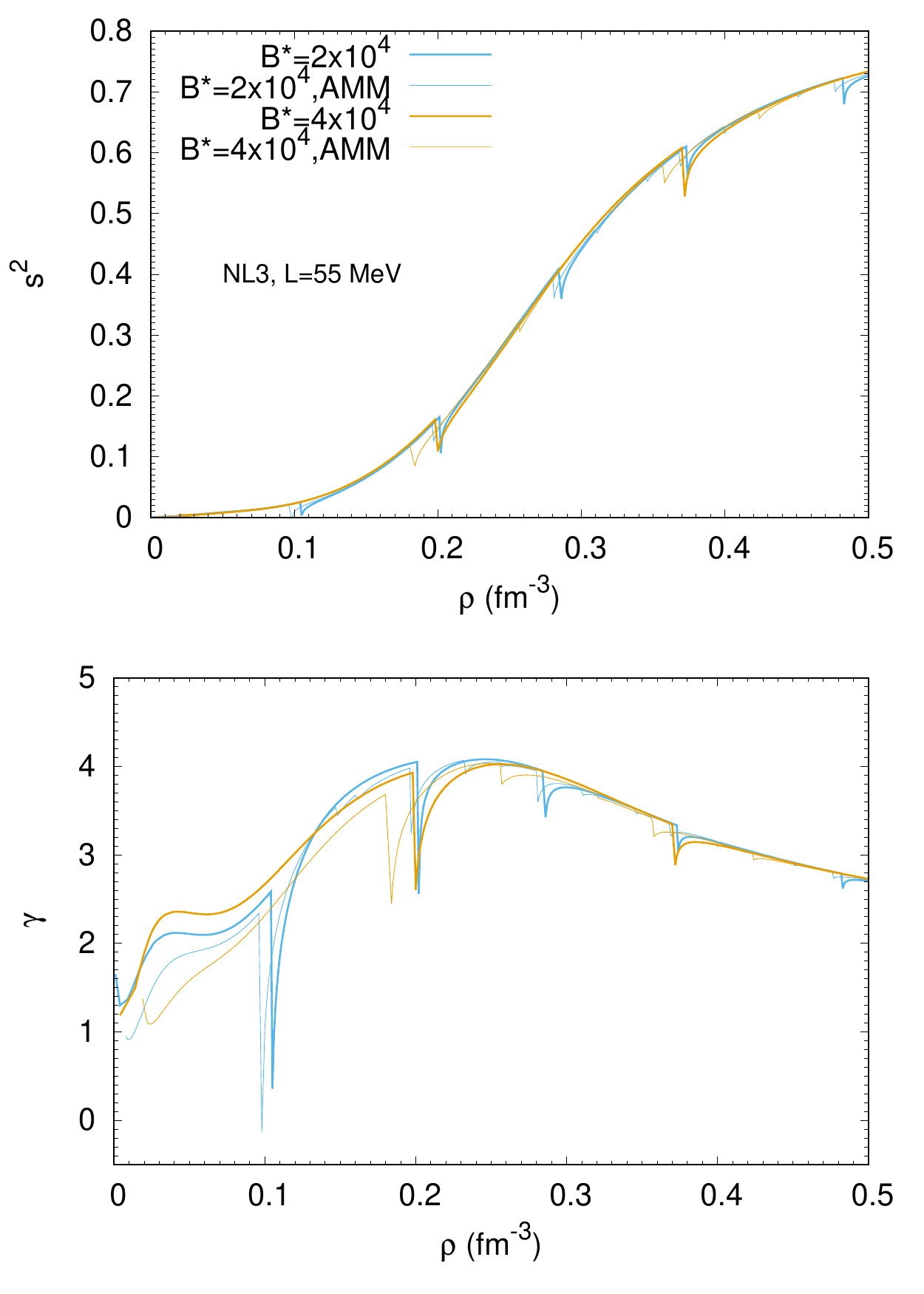}
\end{tabular}
\caption{The square of the sound speed (top), and the adiabatic index (bottom) as a function of the baryonic density $\rho$ calculated for the NL3 model with $L=55$ MeV for $\beta-$equilibrium matter with and without AMM, and different values of the magnetic field.} 
\label{fig3}
\end{figure*}
Fig.~\ref{fig3} shows the speed of sound and the adiabatic index of magnetized nuclear matter in $\beta$-equilibrium calculated within the NL3$\omega\rho$55 model. { In the left panels, we consider AMM=0 and} different values of $B$, $B^*$=0, $10^3$, $2\times 10^4$ and $4\times 10^4$, corresponding to $B \sim 0, 5\times 10^{16}, \, 9\times 10^{17}$, and  $2\times 10^{18}$ G, respectively. In particular, the two last values are quite strong, approximately the strongest fields that are expected inside a NS. The speed of sound and adiabatic index of nuclear matter are quite sensitive to the strength of the applied magnetic field. As it is seen from this Figure, these quantities have a rather complex behavior, due to  the opening of new Landau levels of protons and electrons.
The  speed of sound only shows a visible effect for the two strongest fields of the order of $10^{18}$ G, while  the adiabatic index also shows some small fluctuations due to the opening of Landau levels for $B\sim 5\times 10^{16}$ G, but the overall behavior coincides with the one obtained at $B=0$. Considering the fields $B\sim  10^{18}$ G alone, the opening of a new Landau level may give rise to a sudden reduction of $\gamma$ by one or two units. 

In the right panels of Fig.~\ref{fig3}, we  analyse the effect of the anomalous magnetic moment. The speed of sound and adiabatic index are plotted with (thick lines) and without (thin lines) AMM for the two strongest $B$ fields considered,
because for the weaker fields, the AMM has no  effect. Although, the effect of the AMM is clearly present, no drastic effect is seen and, in the following, we will discuss the effect of the magnetic field on the nuclear matter properties without taking into account the AMM.

\begin{figure}[!]
\centering
   \begin{tabular}{cc}
\includegraphics[width=0.5\columnwidth]{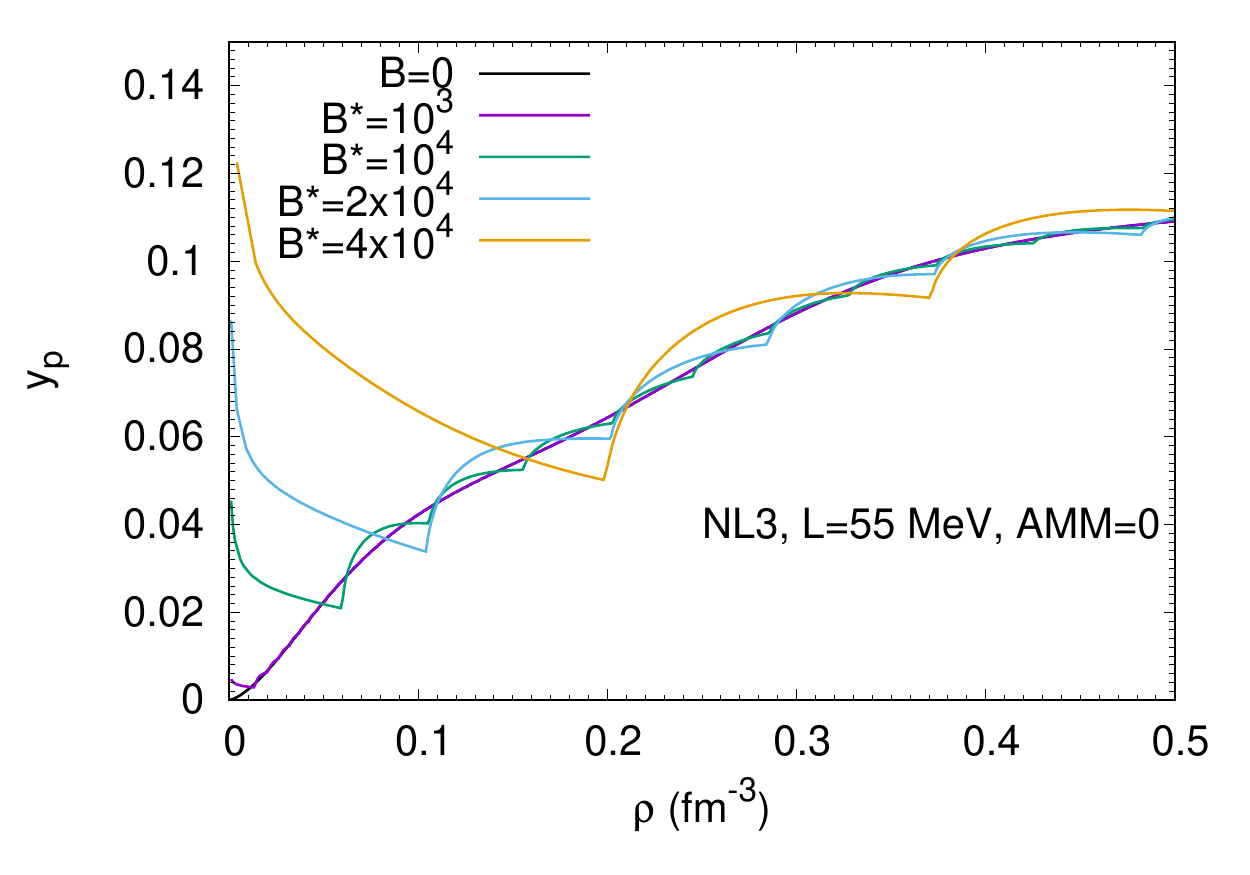} &
\includegraphics[width=0.5\columnwidth]{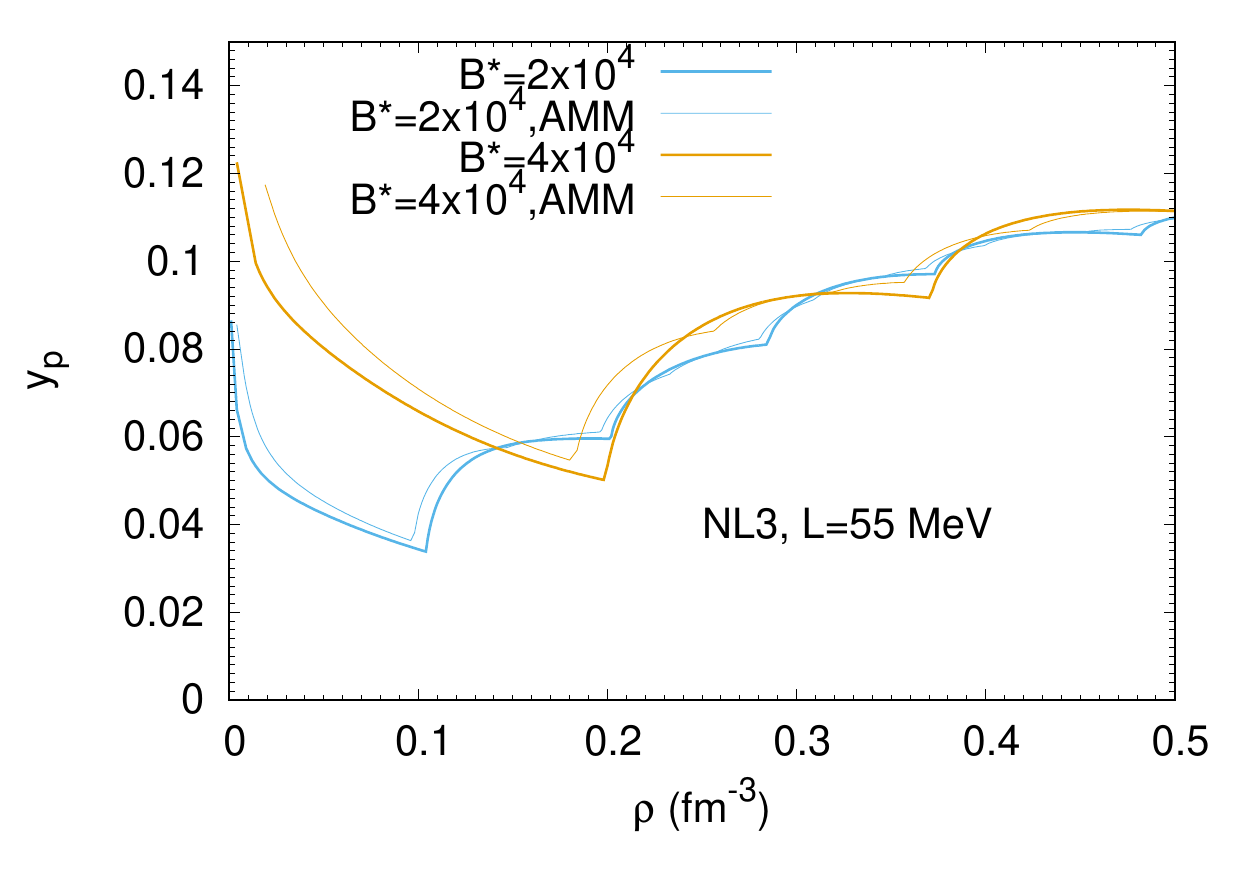} 
     \end{tabular}
\caption{Proton fraction $y_p$ as function of the  baryonic density $\rho$ for cold $\beta$-equilibrium matter with different magnetic field intensities with and without AMM for the NL3 model with $L=55$ MeV.}
\label{fig4a}
\end{figure}

We also would like to pay special attention to the fraction of protons defined as $y_p$. This quantity is especially sensitive to the magnetic field due to the electric charge of protons. 
It is plotted in Fig.~\ref{fig4a}  as a function of $\rho$. The  magnetic field effects are more important at low densities, however, since we are discussing the outer-core nuclear matter properties, we focus on  densities above the crust-core transition $\approx 1/3 \, \rho_0$ to $1/2\, \rho_0$, since below this density, matter is non-homogeneous and requires an adequate description. In the next Section, we will consider densities above 0.05~fm$^{-3}$. The saw-like structure of $y_p$ clearly indicates the opening of a new Landau level. Noticeable effects on the proton fraction in the outer layers of the core occur if the magnetic field is $\gtrsim 5\times 10^{17}$ G. In the right panel of Fig.~\ref{fig4a}, we compare the proton fractions obtained with and without AMM for the two strongest fields and, as before, we conclude that the inclusion of the AMM does not affect much the proton fraction.

\section{Landau parameters and entrainment matrix}
\label{sec4} 

Landau parameters characterize the nuclear interaction at the Fermi surface
and are related with several nuclear matter properties such as the  incompressibility,  the effective nucleon mass and the entrainment matrix. In the following, we will calculate these quantities and discuss how sensitive  they are to the density dependence of the symmetry energy. In particular, we will calculate the parameters $F_{ij}^0$ and  $F_{ij}^1$, and the entrainment matrix $Y_{ij}$  directly related with $F_{ij}^1$.

In fact, the coexistence of the superfluid neutrons and superconducting protons in the neutron star interiors, especially below their crusts, significantly impacts the fluid dynamics. The transportation of momentum of a given component by particles of another one corresponds to the entrainment phenomenon. The physical reason behind the nuclear entrainment is the interaction holding quasi-particles together. Similar correlation between quasi-particles caused by their weak interaction occurs in the helium mixture containing superfluid components of $^3$He and $^4$He. The formalism to treat non-relativistic entrainment in such a system was first developed in Ref. \cite{Andreev1975}.

In this section, the relativistic generalization of the Landau parameters,  discussed in \cite{Caillon2001,Avancini2005,Gusakov2009}, are calculated for all the models we have been discussing. In particular, the entrainment matrix within a RMF approach has been introduced in Ref.~\cite{Gusakov2009,Gusakov2014}, and a detailed description can be found in Ref.~\cite{Pratapsi2017}.  The effect of an external magnetic field on the Landau parameters and  entrainment matrix will also be referred, independently of the possible weakening of superfluidity due to the magnetic field. Indeed, proton superconductivity is likely to be totally removed from the inner core of neutron stars, while being possible in their outer core \cite{Sinha2015}.

 We should also note that our RMF framework has a limited applicability at small densities, where nuclear matter becomes inhomogeneous due to the clusterization effects (see e.g. Ref.~\cite{Pais2018} and references therein). Therefore, { as mentioned before,} we will only discuss properties obtained for densities $\rho > 0.05$ fm$^{-3}$.

The entrainment matrix is directly related with { the Landau parameter} $F_{ij}^1$ as
\begin{equation}
    Y_{ij} = \delta_{ij} \frac{\rho_i}{\mu_i^*}+\frac{1}{3}
    \left(\frac{\rho_i \rho_j}{\mu^*_i \mu^*_j}\right)^2 F_{ij}^1 \, .
    \label{yij}
\end{equation}
It can also be  obtained from the expansion of the nucleon currents $\bf j_n$ and $\bf j_p$  up to linear terms in the corresponding momenta per particle of the Cooper pairs, denoted as $\bf Q_n$ and $\bf Q_p$, i.e.
\begin{eqnarray}
\label{XXVIII}
{\bf j}_n&=&Y_{nn}{\bf Q}_n+Y_{np}{\bf Q}_p,\\
\label{XXIX}
{\bf j}_p&=&Y_{pn}{\bf Q}_n+Y_{pp}{\bf Q}_p.
\end{eqnarray}
The contribution of the normal component is absent in this expansion because we consider a vanishing temperature, which means that all the nucleons exist in the superfluid or superconducting state. The expansion coefficients in Eqs. (\ref{XXVIII}) and (\ref{XXIX}) correspond to the symmetric entrainment matrix
\begin{eqnarray}
\label{XXX}
Y_{nn}&=&\frac{\eta_n(1+\eta_p\Sigma)}
{(1+\eta_n\Sigma)(1+\eta_p\Sigma)-\eta_n\eta_p\Delta^2},\\
\label{XXXI}
Y_{nn}&=&\frac{\eta_p(1+\eta_n\Sigma)}
{(1+\eta_n\Sigma)(1+\eta_p\Sigma)-\eta_n\eta_p\Delta^2},\\
\label{XXXII}
Y_{np}=Y_{pn}&=&-\frac{\eta_n\eta_p\Delta}
{(1+\eta_n\Sigma)(1+\eta_p\Sigma)-\eta_n\eta_p\Delta^2}.
\end{eqnarray}
They are expressed in terms of the ratios $\eta_i={\rho_i}/{\mu_i^*}$ and
\begin{eqnarray}
\label{XXXIII}
\Sigma&=&\left(\frac{g_\omega}{m_\omega^*}\right)^2+
\left(\frac{g_\rho}{2m_\rho^*}\right)^2,\\
\label{XXXIV}
\Delta&=&\left(\frac{g_\omega}{m_\omega^*}\right)^2-
\left(\frac{g_\rho}{2m_\rho^*}\right)^2.
\end{eqnarray}
where $m_\rho^*$ and $m_\omega^*$ are the effective masses of the $\rho$ and $\omega$ mesons, respectively, and they are defined in the Appendix. The entrainment matrix coefficients have dimensionality of ${\rm fm}^{-3}{\rm MeV}^{-1}$. Therefore, and following  Ref. \cite{Gusakov2009}, we introduce the normalization parameter $Y=3.99\cdot10^{-4}~{\rm fm}^{-3}{\rm MeV}^{-1}$, which represents a typical scale.
%

We follow Ref.~\cite{Gusakov2009} to write the first two Landau parameters, $F_0^{ik}$ and $F_1^{ik}$ as
\begin{eqnarray}
F_0^{ii}&=&\sqrt{N_i^2}\left(\Sigma-\frac{g_{\sigma}^2 m_N^{*2}}{L(\sigma)\mu_i^{*2}}\right) \, , \nonumber \\
F_0^{ik}&=&\sqrt{N_iN_k}\left(\Delta-\frac{g_{\sigma}^2 m_N^{*2}}{L(\sigma)\mu_i^{*}\mu_k^*}\right) \, ,  i\neq k \, , \\
F_1^{ik}&=&\sqrt{N_iN_k}\frac{9\pi^4}{k_{Fi}^{2}k_{Fk}^{2}}\left(Y_{ik}-\eta_i\delta_{ik}\right) \, ,  
\end{eqnarray}
where $N_i$ is given by $N_i=\mu_i^* k_{Fi}/\pi^2$ and $L(\sigma)$ by
\begin{eqnarray}
L(\sigma)=m_\sigma^2+k\sigma+\frac{\lambda}{2}\sigma^2+g_\sigma^2\frac{\partial\rho_s}{\partial m^*} \, ,
\end{eqnarray}
and $\rho_s$ is the scalar density, given in the Appendix. Note that under strong magnetic fields, it was argued in Ref.~\cite{PerezGarcia2011} that the density of states $N_p$ should take into account the Landau quantization. { In the present study, we use for the Fermi momentum of the proton the one that corresponds to the first Landau level.}

\begin{figure}
\centering
\includegraphics[width=0.6\textwidth]{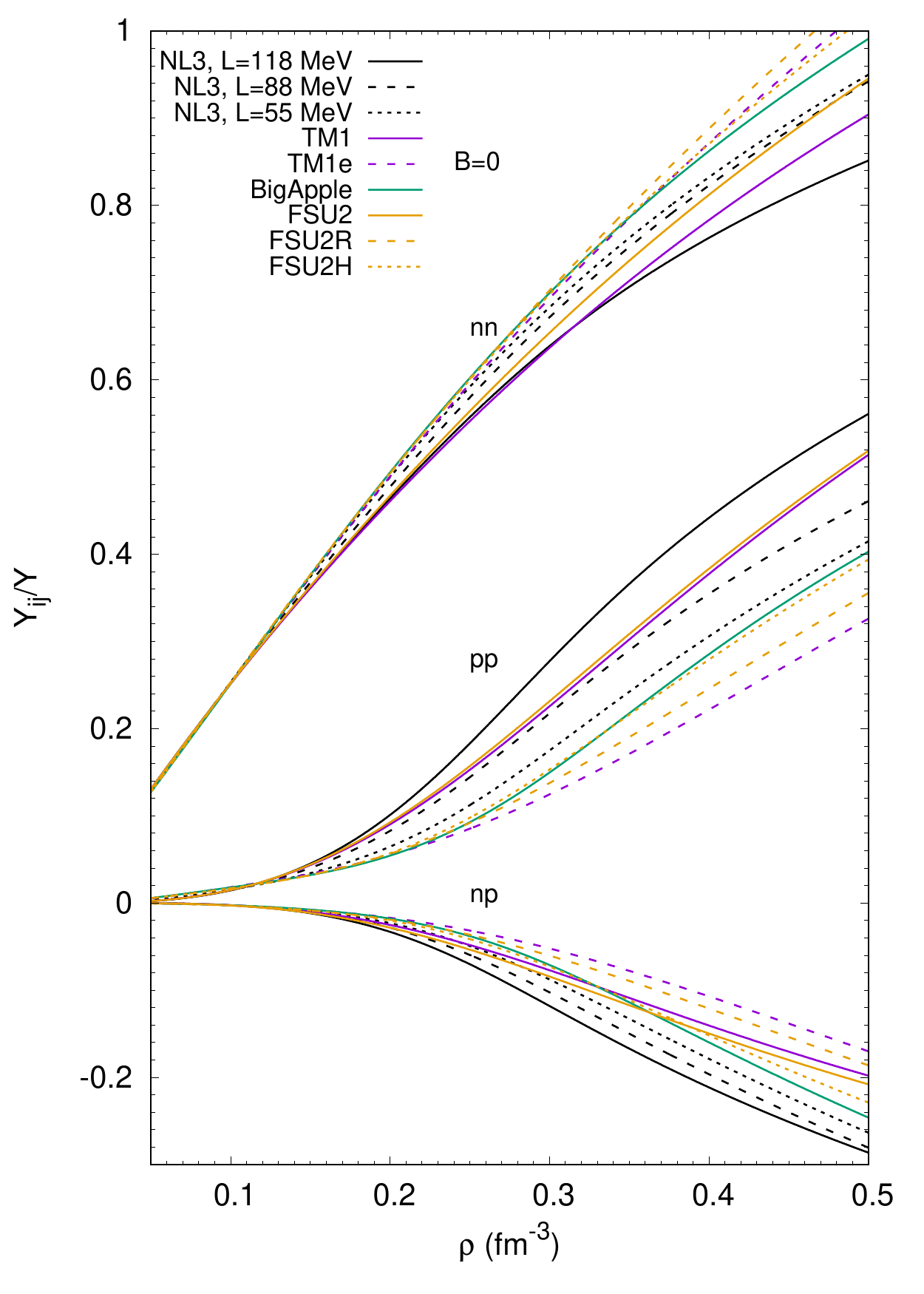} 
\caption{Normalized $Y_{ij}/Y$ entrainment matrix elements as a function of the baryonic density for all the models and cold $\beta$-equilibrium matter without magnetic field. }
\label{fig4}
\end{figure}

\begin{figure*}
\centering
\begin{tabular}{cc}
\includegraphics[width=0.5\textwidth]{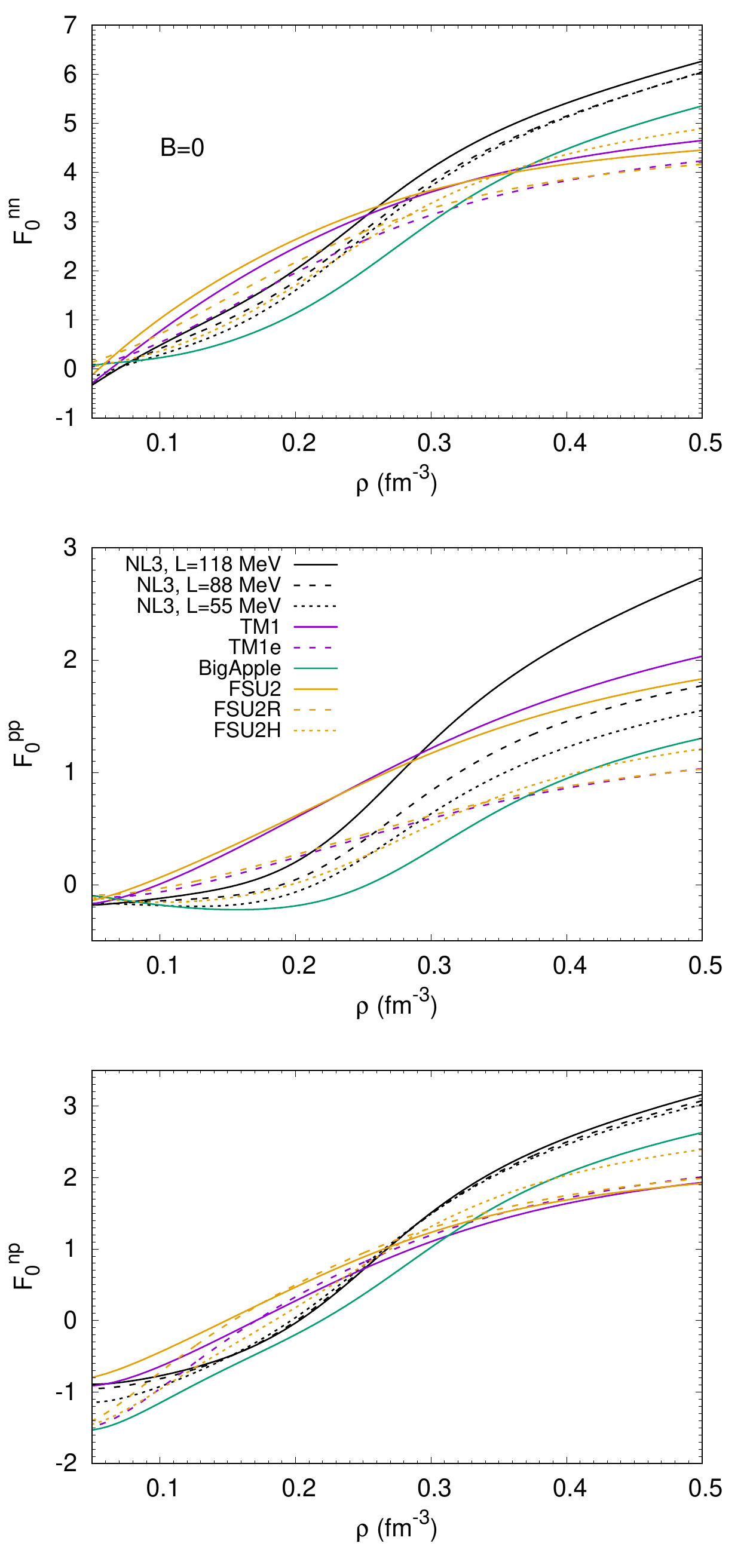} & \includegraphics[width=0.5\textwidth]{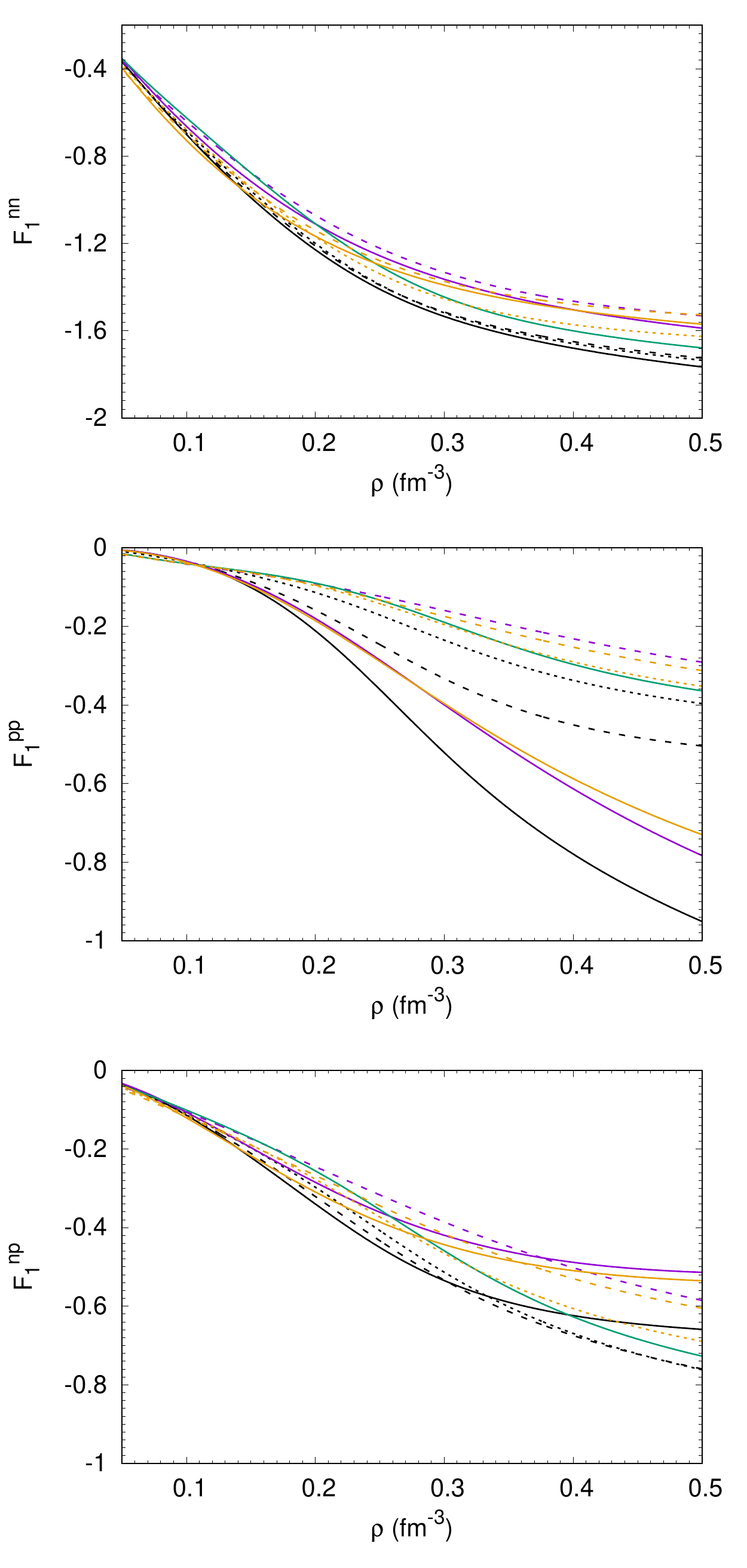}
\end{tabular} 
\caption{$F_0^{ij}$ (left) and $F_1^{ij}$ (right) Landau parameters as a function of the baryonic density for all the models  and cold $\beta$-equilibrium matter without magnetic field. }
\label{fig5}
\end{figure*}

\begin{figure}
\centering
\begin{tabular}{cc}
\includegraphics[width=0.8\textwidth]{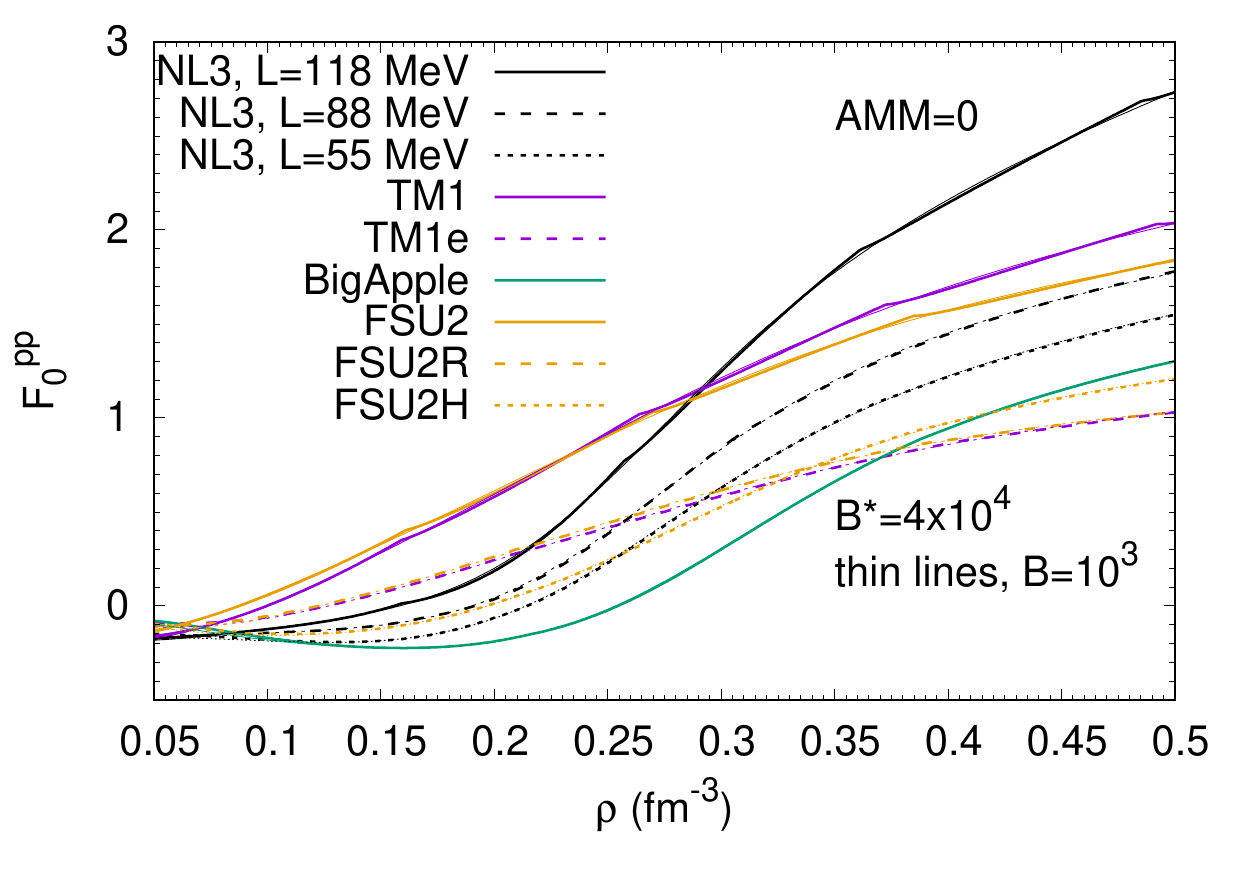} 
\end{tabular} 
\caption{$F_0^{pp}$ Landau parameters as a function of the baryonic density for all the models  and cold $\beta$-equilibrium matter with $B^*=10^3$ (thin lines) and $4\times 10^4$ (thick lines). The AMM is set to zero. }
\label{fig5a}
\end{figure}

We first analyse the impact of the symmetry energy on the   entrainment matrix. In Fig.~\ref{fig4}, the normalized  entrainment matrix elements are shown as a function of the baryonic density for the three sets of  models considered in this work. The positive $Y_{nn}$ and $Y_{pp}$ obviously reflect co-alignment of currents and momenta of nucleons of the same kind, while  the negative value of $Y_{np}$ means that entrainment of nucleons of different species enhances their antiparallel flow.
Moreover, these coefficients demonstrate a clear dependence on the stiffness of the symmetry energy and the value of its slope $L$ at the saturation density. The symmetry energy $J$ describes the energy cost of the difference between the fractions of  the two types of nucleons, and, consequently, a large value disfavors a high population of neutrons compared to protons. A large $L$ leads to the stiffening of the symmetry energy above saturation density, giving rise to less neutrons and more protons in this density regime as compared to a model with a smaller slope $L$. This explains why the increase of $L$ suppresses $Y_{nn}$, and enhances $Y_{pp}$ at $\rho>\rho_0$. { This conclusion can be drawn from this figure} 
for the NL3$\omega\rho$ family, for instance.  $Y_{np}$ is less sensitive to the value of $L$, than $Y_{nn}$ and $Y_{pp}$, since the enhancement of neutrons is partially compensated by the suppression of protons and vice versa. Nevertheless,  the neutron-proton entrainment matrix coefficient is also enhanced by larger values of $L$. The largest values of $Y_{nn}$ are obtained for FSU2R and TM1e, together with FSU2H and BigApple, while the first two show also the smallest $Y_{pp}$ values. In Ref.~\cite{Gusakov2014}, models with a quite large slope $L$ were analysed, and the conclusions drawn are in accordance with the ones obtained in the present work for models with a large symmetry energy slope.

\begin{figure}
\centering
\includegraphics[width=0.7\textwidth]{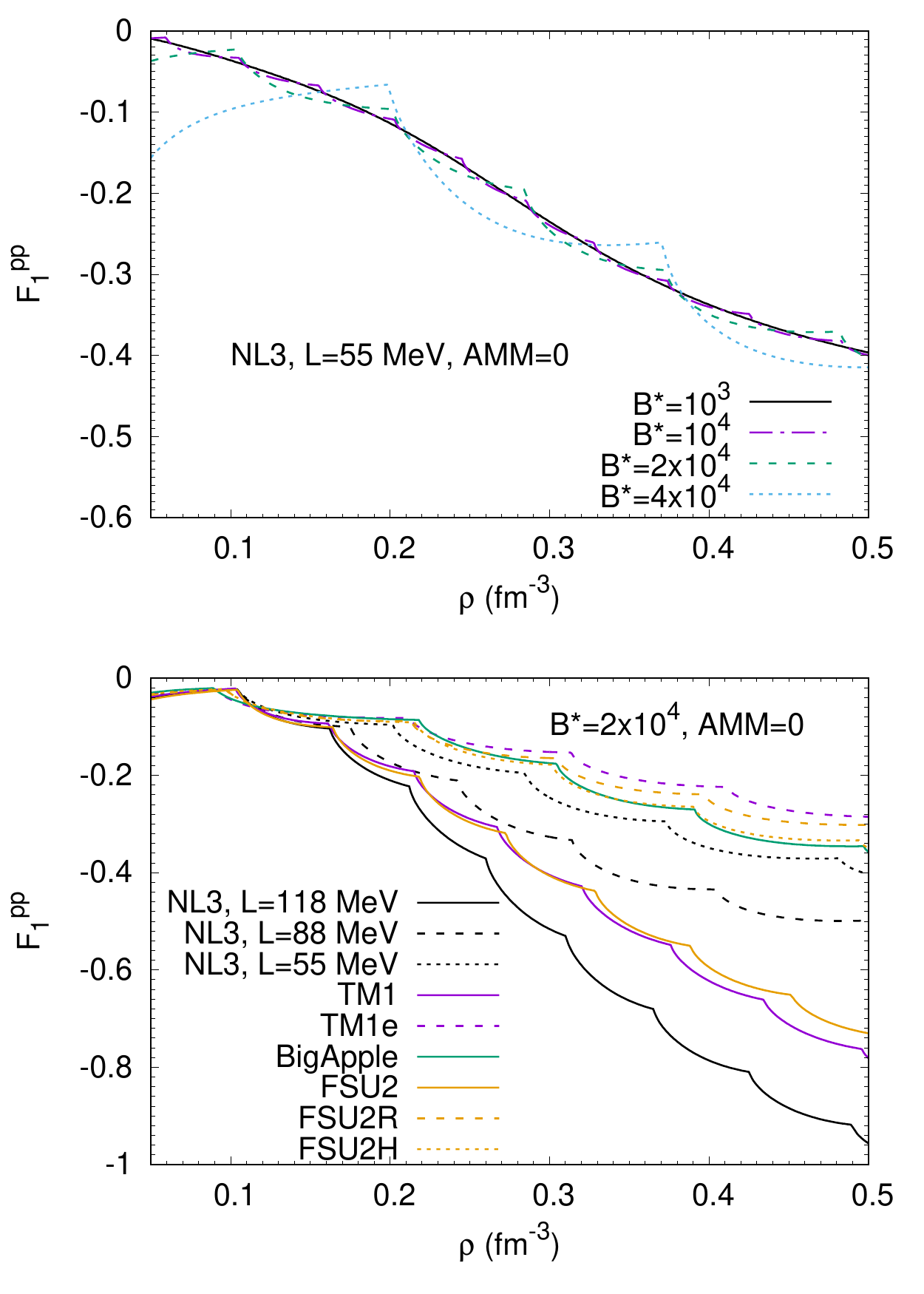}  
\caption{$F_1^{pp}$ Landau parameters  as a function of the baryonic density for cold $\beta$-equilibrium matter, considering the NL3$\omega\rho$55 model and different values of the magnetic field (top panel), and several models and $B^*=2\times 10^4$ (bottom panel). In the top panel, the curve calculated with $B^*=10^3$ coincides with the $B=0$ curve. The anomalous magnetic moment has been set to zero. }
\label{fig7}
\end{figure}

It is interesting to analyse the dispersion of the magnitude of the different entrainment coefficients taking into account the set of models we are studying.
As discussed above, the  neutron-neutron entrainment elements are the largest due to  the high population neutrons. 
Taking all the models, a dispersion of $\approx 0.16$ is obtained on $Y_{nn}$. This results from both the behavior of the symmetry energy and the isoscalar properties. The symmetry energy is responsible for $\gtrsim50\%$ of the dispersion. At $3\rho_0$, the models within the same family differ by 0.08 to 0.1, i.e $\gtrsim 10\%$ of  $Y_{nn}$ at that density. Considering models with a similar symmetry energy behavior, in particular, TM1e, FSU2R, BigApple and NL3$\omega\rho55$, the dispersion on $Y_{nn}$ at $3\rho_0$ is 0.08. This difference is reduced to $\sim 0.05$ if the NL3 family, with a very large incompressibility, is not considered.

Considering now the $Y_{pp}$ coefficients, the overall dispersion is about 0.2 and reduces to 0.15 within a given family, corresponding to $\sim 30\%-50\%$ of the $Y_{pp}$ value. Taking models with a similar symmetry energy slope, the dispersion reduces to $\sim 0.09$ corresponding to about $25\%$ of the magnitude of $Y_{pp}$.
The neutron-proton entrainment matrix coefficients are in absolute value smaller than the $nn$ and $pp$ coefficients. Since the effect of the symmetry energy is opposite in neutrons and protons, a cancellation occurs that reduces the dispersion of the values $Y_{pn}$ to about 0.1. 

\begin{figure}
\centering
\includegraphics[width=0.7\textwidth]{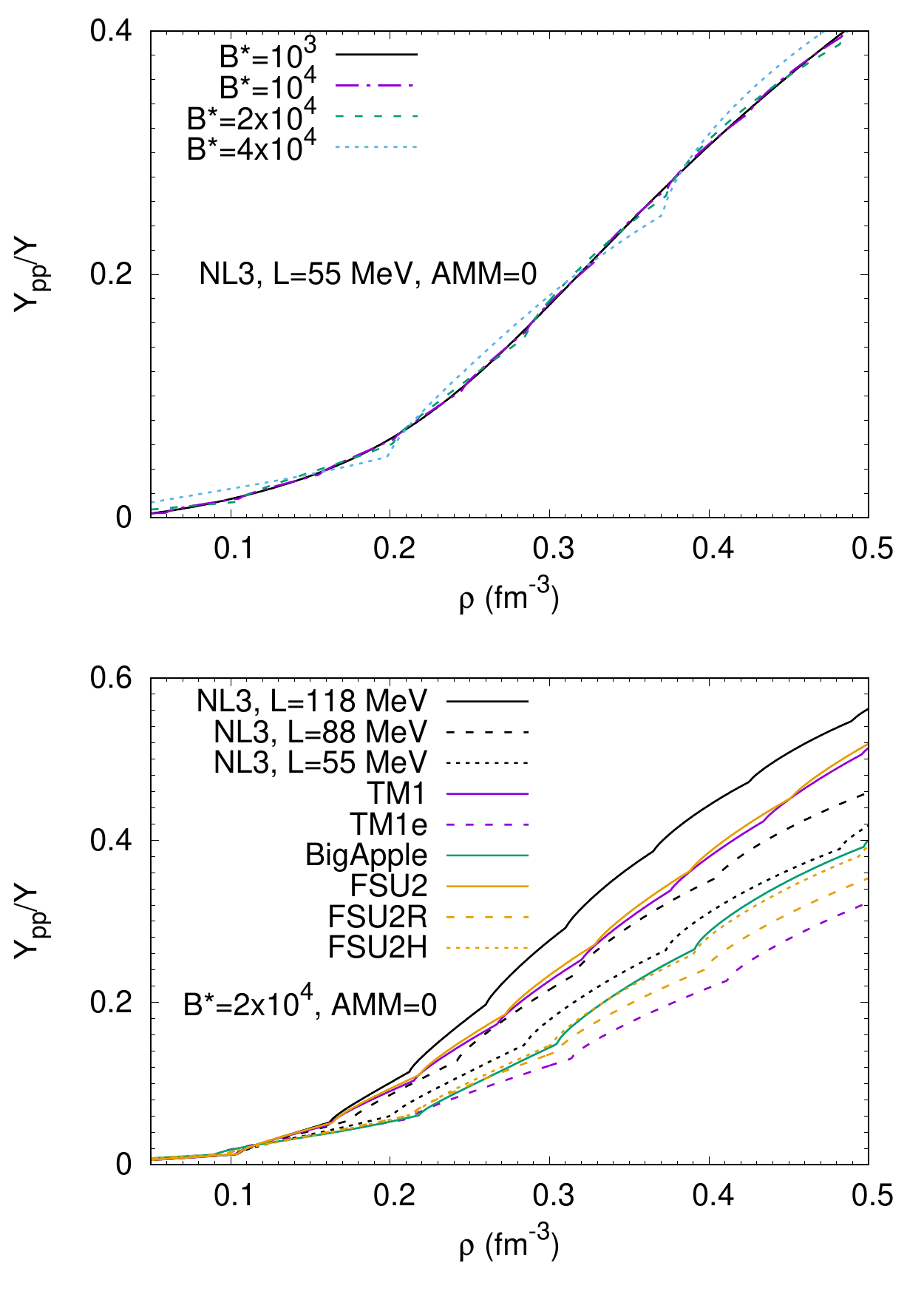}
\caption{Normalized  $Y_{pp}/Y$ entrainment matrix elements  as a function of the baryonic density for cold $\beta$-equilibrium matter, considering the NL3$\omega\rho$55 model and different values of the magnetic field (top panel), and several models and $B^*=2\times 10^4$ (bottom panel). In the top panel, the curve calculated with $B^*=10^3$ coincides with the $B=0$ curve. The anomalous magnetic moment has been set to zero.}
\label{fig7b}
\end{figure}

In Fig.~\ref{fig5}, the Landau parameters $F_l^{ij}$, with $l=0,1$ and $i,j=p,n$, obtained within  all models under study are plotted as a function of density. Stability conditions dictate that \cite{Gusakov2009}
$$ Y_{ii}\ge0 \, , \quad Y_{nn}Y_{pp}-Y_{pn}^2\ge0 \, ,$$ and
$$1+ F_0^{ii}\ge0\quad (1+F_0^{pp})(1+F_0^{nn})-(F_0^{np})\ge0 \, .$$
These conditions are obeyed by all models in the range of densities considered, i.e. for densities above $\approx  0.05$~fm$^{-3}$, which { approximately} defines the crust-core transition. Some comments are in order: (i) the Landau parameters $F_1^{pp}$ may differ by a factor of 100\% or more for models that have the same isoscalar behavior, but differ on the symmetry energy, such as NL3 and NL3$\omega\rho55$, or TM1 and TM1e, or FSU2 and FSU2R, in the range of densities $0.2 - 0.4$ fm$^{-3}$. Taking all the models, it is striking to verify that for $\rho=0.4$ fm$^{-3}$ this parameter may take any value between $-0.2$ and $-0.8$, the more negative values being associated with the larger $L$ slopes; (ii) the dispersion among the values of $F_1^{nn}$ for the same densities and  models  is much smaller, of the order of 12\%; (iii) for $F_1^{np}$,
intermediate values are obtained, and for $\rho=0.4$ fm$^{-3}$, values between $-0.4$ and $-0.6$ are obtained. These differences will certainly reflect themselves on the hydrodynamic behavior of nuclear matter. The dispersion of values obtained for $F_1^{ij}$ was expected, taking into account the behavior of the $Y_{ij}$ coefficients, both quantities being related by Eq.~(\ref{yij}).

Examining the $F_0^{ij}$ parameters, similar conclusions are drawn. The largest differences occur for the $F_0^{pp}$ parameter: the value this parameter takes at $\rho=0.4$ fm$^{-3}$ varies between $\approx1$ and above 2. This is not surprising since they depend on the proton content of matter, and in $\beta$-equilibrium the different models predict quite  different proton fractions.  It is interesting to see that the $F_0^{ij}$ parameters also reflect the  parametrization of the interaction: models with a large $\omega^4$ term have larger values for intermediate densities but increase slower and at  large densities become smaller than the ones with a similar symmetry energy and without the non-linear $\omega$ term.

We conclude this work by showing the effect of the magnetic field on the  Landau parameters $F_0^{pp}$ and $F_1^{pp}$, and on the coefficient $Y_{pp}/Y$. 
 The neutron-neutron parameter is only weakly sensitive to $B$, because it is a neutral particle.  It couples  to $\bf B$ through the AMM, however, stronger fields are necessary to have a noticeable effect. In the following, we will not show results including the AMM, and, therefore, we will restrict our discussion to the proton-proton interaction. The indirect effect on the neutron-neutron parameters due to the Landau quantization experienced by the protons is negligible. 
 
  The Landau quantization of the  proton orbit gives rise to   a series of kinks caused by the opening of new Landau levels, and the Landau parameters  fluctuate around the $B=0$ result.  In Fig.~\ref{fig5a}, the $F^0_{pp}$ parameters are plotted for $B^*=10^3$ and $4\times 10^{4}$, and it is seen that even for a field of the order of $10^{18}$G,  the effects are negligible, and correspond to very soft kinks. The $B^*=10^3$ curve coincides with the $B=0$ curve. 
  
  In the top panel of Fig.~\ref{fig7}, the $F^1_{pp}$ is plotted  for the NL3$\omega\rho$55 model and several field intensities. Non-negligible effects are only obtained for $B^*\gtrsim2\times 10^{4}$. The effect of the magnetic field on $F^1_{pp}$ within the models we have been considering is shown in the bottom panel of Fig.~\ref{fig7} taking $B^*=2\times 10^{4}$.
  Just for reference, the same plots have been done for the coefficients $Y_{pp}$ in Fig. \ref{fig7b}, and the main conclusions are similar.

\section{Conclusions}
\label{sec5}

We have discussed the effect of the symmetry energy on some properties of neutron star matter in $\beta$-equilibrium within a RMF description. In order to analyse the effect,  two models that only differ on the isovector channel where chosen, which we have designated by NL3$\omega\rho$L \cite{NL3fam} and which derive from the NL3 parametrization \cite{NL3}. We have completed the study considering a set of calibrated models recently published, e.g. FSU2R, FSU2H \cite{tolos,tolos2}, TM1e \cite{TM1e}, BigApple \cite{BigApple}, together with the two basic models  FSU2 \cite{fsu2} and TM1 \cite{tm1} from which the four models were obtained. We have calculated properties that determine the EoS and the hydrodynamic behavior of matter inside neutron stars, in particular the Landau parameters and the entrainment matrix, {the Landau effective mass}, the speed of sound and the adiabatic index.

For the models considered, the speed of sound may differ by a factor of 2 for $\rho>0.25$~fm$^{-3}$.
The adiabatic index has shown to be quite sensitive to the density dependence of the symmetry energy between 0.05 and 0.2~fm$^{-3}$, having an approximately common value for $\rho=0.1$~fm$^{-3}$, of the order of 3. At this density, the symmetry energy of all the models considered have a similar value.

Higher values of the symmetry energy slope were found to suppress $Y_{nn}$ and enhance $Y_{pp}$. This is due to the effect of the symmetry energy on the proton fraction: larger values of the slope $L$ give rise to a stiffer symmetry energy above saturation density and, as consequence, larger proton and smaller neutron fractions or a smaller asymmetry between neutrons and protons. It was shown that the isoscalar and isovector properties of the EOS  could have an effect of the order of 10\% on the magnitude of the coefficient $Y_{nn}$, and of 30 to 50\% on the the coefficient $Y_{pp}$. These parameters, and especially $Y_{pp}$, are very sensitive to the density dependence of the symmetry energy that defines the proton fraction of $\beta$-equilibrium matter. Taking the set of models studied, it was shown that the $Y_{pp}$ parameter presents a large  dispersion of values at intermediate densities. This is translated into the Landau parameter $F_1^{pp}$, and for $\rho=0.4$~fm$^{-3}$, this parameter can vary between $-0.2$ and $-0.8$, the lowest values being associated with the models with a larger slope.
The effect on the parameter $Y_{np}$ reflects the compensation of the changes in the population of protons and neutrons.  The parameters $F_0^{ij}$ are more sensitive to the isoscalar behavior of the models, but also depend on the symmetry energy, and, in particular, $F_0^{pp}$ shows a large dependence on the symmetry energy.

We have studied the effect of an external magnetic field on some NS properties. In particular, it was shown that the opening of Landau levels is reflected on the behavior of the speed of sound and adiabatic index, with occurrence of a strong reduction at each opening of a new Landau level. Also the proton fraction of $\beta$-equilibrium matter is strongly affected by the presence of a magnetic field  just above the crust-core transition, which favors larger fractions at low densities: for a field of $\sim 10^{17}$~G, the proton fraction at 0.1 fm$^{-3}$, in the outer core of the NS, may increase to more than the double.
 
 Having focused on densities as the ones occurring in the outer core of a NS, we have concluded that the overall effect of the magnetic field is not so dramatic. The Landau parameters show some kinks due to the opening of the Landau levels if the field is of the order of 10$^{18}$G, but taking the $B=0$ values will give in general a good description of matter. The range of densities more strongly affected is the one just above the crust-core transition and below saturation density. 
 
Finally, we would like to point out that tables with the EoS, and the parameters, for all the models considered in this work at $B=0$, are provided in the Supplementary Material section.

\appendix
\section{Appendix}
\numberwithin{equation}{section}
\label{App}

Performing momentum integration in Eqs. (\ref{XVI}) and (\ref{XVII}), we obtain the single particle contributions to the pressure
\begin{eqnarray}
\label{A1}
P_n&=&\sum_s\left[\frac{
\mu_n^*k_n^F(2\mu_n^{*2}-5(m_n^*-s\mu_N\kappa_nB)^2)
+3(m_n^*-s\mu_N\kappa_nB)^4\ln\frac{\mu_n^*+k_n^F}{m_n^*-s\mu_N\kappa_nB}}{48\pi^2}\right.\nonumber\\
&+&s\mu_N\kappa_nB\left.\frac{
2\mu_n^*k_n^F(m_n^*-s\mu_N\kappa_nB)
-\mu_n^{*3}{\rm arcsin}\frac{k_n^F}{\mu_n^*}
-(m_n^*-s\mu_N\kappa_nB)^3\ln\frac{\mu_n^*+k_n^F}{m_n^*-s\mu_N\kappa_nB}}{12\pi^2}\right],\\
\label{A2}
P_p&=&\sum_\nu g_\nu\frac{eB}{4\pi^2}
\left(\mu_p^*k_{||p}^F-\left(\sqrt{2\nu eB+{m_N^*}^2}-s\mu_N\kappa_pB\right)^2
\ln\frac{\mu_p^*+k^F_{||p}}{\sqrt{2\nu eB+{m_N^*}^2}-s\mu_N\kappa_pB}
\right),\\
\label{A3}
P_e&=&\sum_\nu g_\nu\frac{eB}{4\pi^2}
\left(\mu_e k_{||e}^F-(2\nu eB+m_e^2)
\ln\frac{\mu_e+k^F_{||e}}{\sqrt{2\nu eB+m_e^2}}
\right),
\end{eqnarray}
where $k_n^F=\sqrt{\mu_n^{*2}-(m_n^*-s\mu_N\kappa_nB)^2}$, and the other quantities were already introduced in Section \ref{sec3}.

 Solving the Euler-Lagrange equations, and in the mean-field approximation, the equations for the fields read:

\begin{eqnarray}
\label{A4}
{m_\sigma^*}^2\sigma\phantom{^0}&=&\gs\rho_s=\gs 
({\rho_s^n +\rho_s^p}) \\
\label{A5}
{m_\omega^*}^2\omega^0 &=& \gw\rho=\gw (\rho_n + \rho_p)  \\
\label{A6}
{m_\omega^*}^2{\boldsymbol \omega}\phantom{^0} &=& \gw ({\bf j}_p + {\bf j}_n)  \\
\label{A7}
{m_\rho^*}^2 b_0 &=& \frac{\gr}{2}\rho_3 =\frac{\gr}{2}  (\rho_p - \rho_n)\\
\label{A8}
{m_\rho^*}^2{\boldsymbol b}^{\phantom{0}} &=& \frac{\gr}{2}({\bf j}_p - {\bf j}_n),
\end{eqnarray}
with the meson effective masses $m_i^*$ defined as:
\begin{eqnarray}
\label{A9}
{m_\sigma^* }^2 &=& m_\sigma^2 + \frac{\kappa \sigma}{2} + \frac{\lambda \sigma^2}{6} \\
\label{A10}
{m_\omega^*}^2 &=& m_\omega^2 + \frac{\zeta}{6}\gw^4\omega_0^2
+ 2 \gwr \, \gw^2 \, \gr^2 \,b_0^2 \\
\label{A11}
{m_\rho^*}^2 &=& m_\rho^2 + 2 \gwr \, \gw^2 \, \gr^2 \,\omega_0^2.
\end{eqnarray}
The particle number densities are given by
\begin{eqnarray}
\label{A12}
\rho_n&=&\sum_s\left[\frac{k_n^{F3}}{6\pi^2}+
\frac{s\mu_N\kappa_n B}{4\pi^2}\right.\nonumber\\
&\times&\left.
\left(\mu_n^{*2}{\rm arcsin}\frac{k_n^F}{\mu_n^*}-(m_n^*-s\mu_N\kappa_n B)k_n^F\right)\right],\quad\\
\label{A13}
\rho_i&=&
\sum_\nu g_\nu \frac{e B}{2\pi^2}k_{||i}^F,\quad i=p,~e
\end{eqnarray}
and the scalar densities are computed from
\begin{eqnarray}
\label{A14}
\rho_s^n&=&\sum_s\frac{m_n^*}{4\pi^2}\biggl[\mu_n^*k_n^F\nonumber\\
&-&\left.
(m_n^*-s\mu_N\kappa_nB)^2\ln\frac{\mu_n^*+k_n^F}{m_n^*-s\mu_N\kappa_nB}\right] \, , \\
\label{A15}
\rho_s^p&=&\sum_\nu g_\nu\frac{eB}{2\pi^2}
\frac{\left(\sqrt{2\nu eB+{m_N^*}^2}-s\mu_N\kappa_pB\right)m_N^*}{\sqrt{2\nu eB+{m_N^*}^2}}\nonumber\\
&\times&\ln\frac{\mu_p^*+k^F_{||p}}
{\sqrt{2\nu eB+{m_N^*}^2}-s\mu_N\kappa_pB}.
\end{eqnarray}
The proton and neutron currents are given by
\begin{eqnarray}
\label{A16}
\bm j_n &=& \frac{\rho_n}{\mu_n^*}(\Qn - \gw \bm \omega + \frac{\gr}{2}\bm b) \, , \\
\label{A17}
\bm j_p &=& \frac{\rho_p}{\mu_p^*}(\bm \Qp - \gw \bm \omega - \frac{\gr}{2}\bm b),
\end{eqnarray}
where $\bf Q_n$ and $\bf Q_p$ are momenta per particle of the Cooper pairs of neutrons and protons, respectively.


\acknowledgments
This work was partly supported by the FCT (Portugal) under the Projects No. UID/\-FIS/\-04564/\-2019, No. UIDP/\-04564/\-2020, No. UIDB/\-04564/\-2020, and No. POCI-01-0145-FEDER-029912 with financial support from Science, Technology and Innovation, in its FEDER component, and by the FCT/MCTES budget through national funds (OE), and by PHAROS COST Action CA16214. H.P. acknowledges the grant CEECIND/03092/2017 (FCT, Portugal).


\end{document}